\documentclass[%
 aip, pof, preprint,graphics,graphicx,longbibliography,
 amsmath,amssymb,
author-year%
]{revtex4-1}

\usepackage{graphicx}
\usepackage{dcolumn}
\usepackage{bm}
\usepackage[mathlines]{lineno}

\usepackage[utf8]{inputenc}
\usepackage[T1]{fontenc}
\usepackage{mathptmx}
\usepackage{etoolbox}
\usepackage{xcolor,amsmath,amsthm,indentfirst}
\usepackage[shortlabels]{enumitem}

\makeatletter
\def\@email#1#2{%
 \endgroup
 \patchcmd{\titleblock@produce}
  {\frontmatter@RRAPformat}
  {\frontmatter@RRAPformat{\produce@RRAP{*#1\href{mailto:#2}{#2}}}\frontmatter@RRAPformat}
  {}{}
}%
\makeatother
\begin{document}

\title[Clusters of turbulent reattachment]{Evolution of clusters of turbulent reattachment due to shear layer instability in flow
past a circular cylinder}
\author{Gaurav Chopra} 
 \affiliation{Department of Aerospace Engineering, Indian Institute of Technology Madras, 600036 Chennai, India}
 \affiliation{Centre for Excellence for studying Critical Transitions in Complex Systems, Indian Institute of Technology Madras, 600036 Chennai, India}
\author{Sanjay Mittal} 
 \homepage{https://home.iitk.ac.in/~smittal/}
\affiliation{Department of Aerospace Engineering, Indian Institute of Technology Kanpur, 208016 Kanpur, India}%

\author{R.I. Sujith}
 \email{sujith@iitm.ac.in}
\homepage{http://www.ae.iitm.ac.in/~sujith/}
\affiliation{Department of Aerospace Engineering, Indian Institute of Technology Madras, 600036 Chennai, India}
\affiliation{Centre for Excellence for studying Critical Transitions in Complex Systems, Indian Institute of Technology Madras, 600036 Chennai, India}

\date{\today}

\begin{abstract}
We perform large eddy simulations of flow past a circular
cylinder for the Reynolds number ($Re$) range, $2\times 10^3 \leq Re \leq 4\times10^5$, spanning subcritical, critical and supercritical regimes.
We investigate the spanwise coherence of the flow in the critical and supercritical regimes using complex networks. 
In these regimes, the separated flow reattaches to the 
surface in a turbulent state due to the \textcolor{black}{turbulence generated by the} shear layer instability. In the early critical regime, the turbulent reattachment does 
not occur simultaneously at all span locations.
It occurs incoherently along the span in clusters.  
We treat strong surface pressure fluctuations due to the shear layer instability as extreme events and construct time-varying spatial proximity networks where links are based on synchronization between events. This analysis unravels the underlying complex spatio-temporal dynamics by enabling the estimation of characteristics of clusters of turbulent reattachment via the concept of connected components.
In the critical regime, the number and size of the clusters increase with increase in $Re$. At higher $Re$ in the supercritical regime, they coalesce to form bigger clusters, resulting in increase in spanwise coherence of turbulent reattachment. We find that the size and number of clusters govern the variation of the time-averaged coefficient of drag ($\overline{C}_D$) in the critical and supercritical regimes. \textcolor{black}{$\overline{C}_D$ exhibits power-law distribution with the largest cluster size 
($\overline{C}_D \propto {\overline{S}_{CL}}^{-\frac{2}{5}}$)  and the most probable cluster size 
($\overline{C}_D \propto E(S_C)^{-\frac{2}{5}}$) .
}

\end{abstract}

\maketitle

\section{\label{sec:introduction} Introduction \protect}

\begin{figure}[b]
	\centerline
	{\includegraphics[width=0.7\textwidth]{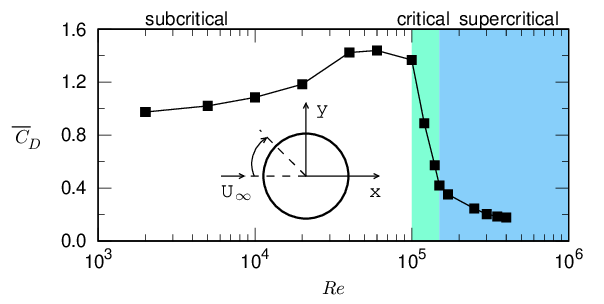}}
	\caption{Flow past a circular cylinder: variation of time-averaged coefficient of drag ($\overline{C}_D$) with $Re$.  The inset shows a schematic of the problem setup.}
\label{fig:intro_figure}
\end{figure}

The flow past a circular cylinder is classified into subcritical, critical, and supercritical regimes based on
 the state of the boundary layer
 \citep{roshko1961experiments,achenbach1968distribution}. The state depends on the Reynolds number ($Re$), which is defined as 
 $Re={{\rho} U_{\infty} D}/{\mu}$, where $\rho$ is the fluid density, $U_{\infty}$ is the free stream velocity, $D$ is the diameter of the cylinder and $\mu$ is the dynamic viscosity of the fluid. Various flow regimes for a cylinder with span length, $L_z=1D$, are marked in figure \ref{fig:intro_figure} that shows the variation of the time-averaged coefficient of drag ($\overline{C}_D$) with $Re$.  
 This data is obtained via large eddy simulation (LES) and reported in our earlier studies \citep{chopra2022effect,chopra2021jfm}.

 In the subcritical regime, the boundary layer
 separates from the shoulder in a laminar state.
 The separated shear layer in the wake rolls-up 
 into smaller shear layer vortices via the Kelvin-Helmholtz mode of instability
 \citep{williamson1996vortex}.
 As $Re$ increases, the location of the shear layer instability moves 
 upstream 
\citep{prasad1997instability,singh2005flow}.
The sub-critical regime is followed by the critical regime, wherein 
the shear layer instability occurs very close to the surface, leading to flow reattachment
\citep{singh2005flow,kim_choi_park_yoo_2014}. 
The reattached flow develops into a turbulent boundary layer which 
separates at a higher angle. The turbulent flow separation point moves 
downstream with an increase in $Re$, narrowing the wake \citep{chopra2021jfm}.
This results in a sharp decrease in $\overline{C}_D$ with increase in $Re$. 
Therefore, this phenomenon is 
popularly known as the drag crisis 
\citep{landau1982fluid}. Separation and reattachment of the 
flow lead to the formation of a laminar separation bubble (LSB).
The critical regime is followed by
the supercritical regime where the drop in $\overline{C}_D$ with $Re$ is 
relatively gradual compared to that in the critical regime.

\textcolor{black}{
In the present work, we study the spatio-temporal behavior of the shear layer instability that leads to turbulent reattachment and formation of LSB in flow past a circular cylinder via large eddy simulations (LES). Recently, \citet{desai2022effect} explored the evolution of the LSB in the critical regime of flow past a smooth sphere using wind tunnel experiments. They reported that along with intermittency in time, the LSB also exhibits fragmentation in space. In the lower end of the critical regime, the LSB does not appear simultaneously at all the azimuthal locations. It appears in fragments/clusters at some azimuthal locations, leading to non-axisymmetric flow reattachment. As the $Re$ increases, the LSB develops at more azimuthal locations while exhibiting intermittency in time \citep{deshpande2017intermittency}. Finally, at the end of the critical regime, the LSB appears at all the azimuthal locations, leading to an axisymmetric flow. Such a spatio-temporal analysis of the topology of LSB has not been performed for flow past a circular cylinder that includes the spanwise variation. Earlier studies explored temporal intermittency of the LSB either for flow at a single span station \citep{cadot2015statistics} or for span-averaged flow \citep{chopra2017intermittent}, thereby neglecting the dynamics of the LSB along the span of the cylinder.}

\par
\begin{figure}
	\centerline
	{\includegraphics[width=0.7\textwidth]{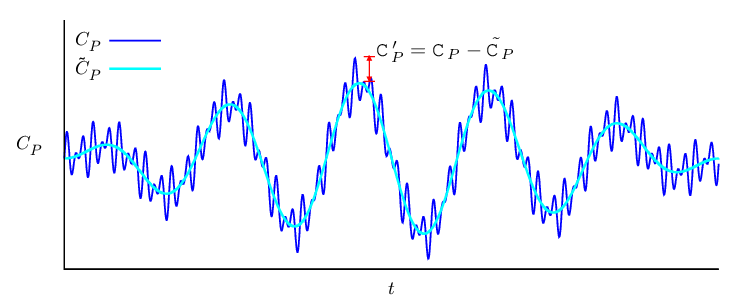}}
	\caption{Flow past a circular cylinder: schematics of times-series of the surface pressure coefficient ($C_P$) and moving averaged $C_P$ ($\widetilde{C}_P$) in the critical and supercritical regimes.}
\label{fig:intro_figure_schematic}
\end{figure}

 \textcolor{black}{Flow past a circular cylinder in the critical and supercritical regimes is associated with three distinct time scales \citep{chopra2017intermittent,desai2020vortexshedding,chopra2021jfm}. \textcolor{black}{ These time scales are described in figure \ref{fig:intro_figure_schematic} via a schematic of a time series of the coefficient of pressure ($C_P(z,\theta,t)$) at a point on the cylinder surface (defined using spanwise ($z$) and azimuthal ($\theta$) coordinates).} The smallest time scale is associated with high-frequency variations due to vortices generated by the shear layer instability in close proximity to the surface. This activity is intermittent in the critical regime, resulting in intermittent turbulent reattachment and formation of LSB \citep{cadot2015statistics,chopra2017intermittent,deshpande2017intermittency}. The second time scale is concerned with the activity due to the periodic von K\'arm\'an vortex shedding, and the largest one is the low-frequency amplitude modulation driven by the expansion/contraction of the vortex formation region  \citep{desai2020vortexshedding,forouzi2022review}.}

        \textcolor{black}{
Based on this observation, \citet{chopra2017intermittent} proposed a double decomposition of surface pressure coefficient: $C_{P}(z,\theta,t)=\widetilde{C}_P(z,\theta,t)+C'_P(z,\theta,t)$,
where $C'_P$ is the contribution from activity due to shear 
layer instability, and $\widetilde{C}_P$ is the contribution from vortex shedding 
and expansion/contraction of the vortex formation region (see 
figure \ref{fig:intro_figure_schematic}).
$\widetilde{C}_P$ is estimated by performing a moving average of $C_P$ over a 
few cycles of shear layer instability: 
$\widetilde{C}_P(z,\theta,t)=\frac{1}{T} \int_{t-T/2}^{t+T/2} \! 
C_P(z,\theta,t) \, \mathrm{d}t$. 
\citet{chopra2017intermittent} proposed the window size to be 
one-tenth of the time period for the vortex shedding cycle ($T_k$), i.e., $T=T_k/10$. 
As seen in figure 2, 
the activity due to the shear layer instability is suppressed in $\widetilde{C}_P$. 
Therefore, one can consider $\widetilde{C}_P$ as low pass filtered $C_P$.
Subsequently, \citet{chopra2017intermittent} estimated fluctuations due to the shear layer instability using: $C'_P(z,\theta,t)=C_P(z,\theta,t)-\widetilde{C}_P(z,\theta,t)$.
\citet{chopra2017intermittent} showed that the moving averaged
$rms$ of $C'_P$ ($=\widetilde{C}'_{P~{rms}}$) is a good measure for 
tracking instances of shear layer instability leading to 
turbulent reattachment and formation of LSB.
They further utilized this approach to estimate 
the intermittency of transition due to the shear layer instability 
generated by a boundary layer trip \citep{chopra2022effect}.
        }


\textcolor{black}{We conduct the spatio-temporal analysis of the surface pressure fluctuations due to the shear layer instability ($\widetilde{C}'_{P~{rms}}$) that lead to turbulent reattachment using a novel complex network approach \citep{barabasi2016network}.} Complex network theory has emerged as a promising tool to analyze turbulent spatio-temporal flow fields \citep{iacobello2021review,taira2022network}. \textcolor{black}{Turbulent flows are characterized by enhanced values of fluid properties such as mixing, diffusion, dissipation, and momentum flux, occurring in intermittent bursts. Bursts are large amplitude fluctuations over a very short time duration. They are associated with the formation and convection of coherent structures popularly referred to as eddies. Consequently, \citet{narasimha1990turbulent} and \citet{narsimha1995turbulence} suggested treating eddies and bursts as episodic extreme events. According to \citet{narsimha1995turbulence}, such description is more natural than the spectral description wherein eddies are treated as waves. Moreover, this approach can help gain further insights into the dynamics of turbulent flows than what is already known from the spectral description. \citet{arakeri2021turbulence} while discussing \citet{narsimha1995turbulence} proposed that new flow analysis methods should be developed that are able to incorporate the event description of eddies. \citet{arakeri2021turbulence} claimed that such methods could potentially improve turbulence models for simulating flow past complex geometries and atmospheric flows. In the present study, we build upon this notion and propose a methodology based on complex network based analysis that incorporates the event description of the turbulent fluctuations. We consider intermittent strong fluctuations in surface pressure due to the shear layer instability as extreme events. We show that this perspective unravels hidden physical mechanisms associated with the transition of boundary layer and drag-crisis in flow past a circular cylinder.}  

\textcolor{black}{Complex network analysis involves embedding the flow field into a graph or network consisting of nodes and links \citep{iacobello2021review,taira2022network}. In the present study, nodes are discrete grid points of the computational mesh. However, for cases where data is obtained via experiments, nodes can be the spatial points in the domain where measurements are made. Links represent interactions between the nodes. Unlike in the power grid or social networks, links in complex network based on fluid flows are not physical; rather, they are abstract. Thus, we rely on statistical similarity in fluid flow properties between two nodes to quantify interactions and establish links. For spatio-temporal analysis of flow fields, researchers have proposed several measures to quantify interactions and establish links such as velocity induced by vorticity in vorticity networks \citep{taira2016network} and correlation between flow variables \citep{scarsoglio2016complex,unni2018emergence} among many others.}
 
\textcolor{black}{\citet{taira2016network} discovered that the vorticity network for two-dimensional 
decaying isotropic turbulence has a scale-free nature, i.e., the
flow field contains a broad range of vortical interactions in terms of their strength.
They showed that regions of strong vortical 
interactions, referred to as hubs, are rare and targeted action
on them, such as their removal, can drastically change the flow field.
Later, \citet{krishnan2021suppression} showed via experiments in a practical 
turbulent combustor that, indeed, the action of
flow control strategies on hubs in the vorticity network can suppress high-amplitude 
oscillations associated with thermoacoustic instability.
\citet{meena2018network} using community detection in vortical networks, developed 
reduced-order models for flow past airfoils and bluff bodies.
Using a similar approach, \citet{meena2021identifying} characterized coherent 
structures in a three-dimensional isotropic turbulent flow.
\citet{iacobello2021large} demonstrated the effectiveness of using a 
visibility-network based approach to study frequency modulation due to inter-scale 
interactions in a turbulent boundary layer. \textcolor{black}{Recently, \citet{iacobello2022load} used a modified cluster-based transition networks approach \citep{doi:10.1126/sciadv.abf5006} to estimate aerodynamic loads from sparse and noisy pressure measurements obtained from experiments.}
Interested readers are encouraged to read
reviews by \citet{iacobello2021review} and \citet{taira2022network} on applying 
complex network theory for fluid dynamics research.}

\textcolor{black}{In the present study, we establish links when the events of strong fluctuations due to the shear layer instability (\textcolor{black}{$\widetilde{C}'_{P~{rms}}$}) occur simultaneously on two nodes that are in proximity of each other. A similar network construction method was used by \citet{krishnan2019emergence} for studying the evolution of clusters of acoustic power sources in turbulent combustors. This network is often referred to as a spatial proximity network \citep{iacobello2021review,rokach2010data}. The spatial proximity network used here is based on the Eulerian perspective of the flowfield. Researchers have also constructed spatial proximity networks from the Lagrangian perspective, wherein they establish links based on the proximity of trajectories of infinitesimal fluid elements \citep{PhysRevE.93.063107,npg-24-661-2017,macau2018mathematical,iacobello_scarsoglio_kuerten_ridolfi_2019,iacobello2021review}. Combined with clustering methods this network approach is often utilized for performing flow classification in conventional and geophysical fluid flows, especially for identifying Lagrangian coherent vortices/structures \citep{PhysRevE.93.063107,npg-24-661-2017}. Spatial proximity networks from the Lagrangian perspective have been used to identify regions of strong mixing in turbulent channel flow \citep{iacobello_scarsoglio_kuerten_ridolfi_2019} and recently, by \citet{shri2022complex} to study the clustering of inertial particles in particle-laden Taylor-Green flow.}
        
We show that the turbulent reattachment due to shear layer instability does not occur uniformly across the cylinder span in the early critical regime. Rather, it occurs in isolated clusters or islands along the span of the cylinder, similar to what \citet{desai2022effect} observed in flow past a sphere. The spatial proximity network analysis enables the identification of clusters of turbulent reattachment, studying their evolution with $Re$, and discovering the intricate physics involving growth in the number of clusters and their coalescing. Further, we show that the evolution of cluster characteristics \textcolor{black}{is associated with} the variation of $\overline{C}_D$ in the critical and supercritical regimes.

\section{Computational details}
We perform LES for flow past a circular cylinder for the 
$Re$ range, $2\times10^3 \leq Re \leq 4\times10^5$.
Results shown in this work are obtained from the same computational setup
as in our recent studies, \citet{chopra2021jfm,chopra2022effect,chopra2023laminar}.
The simulations are performed by solving the governing
equations for incompressible flow using a 
stabilized finite element method \citep{tezduyar1992incompressible}.
Stabilization against possible numerical oscillations is achieved using the
streamline-upwind/Petrov-Galerkin and pressure-stabilizing/Petrov-
Galerkin methods \citep{tezduyar1992incompressible}.
Spatial discretization is performed using elements with equal-order interpolation for velocity and pressure. We use the Crank-Nicholson scheme with
second-order accuracy for time integration. The algebraic equations
resulting from the discretization are solved using the matrix-free generalized
minimal residual method with diagonal preconditioners 
We use the sigma sub-grid scale model 
\citep{nicoud2011using} to account for the effects of the unresolved 
sub-grid scales in the flow.

The governing equations are solved in a hexagonal computational domain.
A cylinder with diameter $D$ and $L_z=1D$ spans the entire domain along the 
$z-$axis. The ``no-slip" boundary condition on the velocity is applied on the 
surface of the cylinder. The ``slip-wall" boundary condition is prescribed on
the lateral, upper, and lower boundaries. Uniform flow is assigned at the inlet 
boundary, while the stress vector is specified as zero on the outflow boundary.
The length of the computational domain in the streamwise direction ($x-$ axis) is 
$38D$, and $16D$ in the cross-stream direction ($y-$ axis).
We use an unstructured mesh with approximately 
$6\times10^6$ nodes and $12\times10^6$ elements. The three-dimensional mesh is
generated by stacking several copies of two-dimensional mesh along the span
with a spacing of $\Delta z=0.02D$. The number of nodes on the 
surface along the circumferential direction in the two-dimensional mesh is 
$N_{\theta}=800$. The height of the elements on the surface 
is $5\times10^{-6} D$. To adequately resolve all the time scales in the flow,
we progressively decrease the non-dimensional time step from 
$\Delta t=5\times10^{-4}$ at the onset of the critical regime to 
$5\times10^{-5}$ at $Re=4\times10^5$ in the supercritical regime, which is the
highest $Re$ considered in this study. Time is non-dimensionalized with $D/U_\infty$.

The results from present computations are in very good agreement with
those from earlier studies.
A detailed validation study of the present computations against well-established 
experiments and computations across all regimes were carried out and reported in 
our earlier study \citep{chopra2021jfm}.
We performed several tests to check the 
adequacy of the grid size resolution.
We estimate the $y^+={yv^*}/{\nu}$ corresponding to the 
height of the first element on the surface of the cylinder at $Re=4\times10^5$.
Here $y$ is the distance of the field node from the cylinder surface, $\nu$ is 
the kinematic viscosity, and $v^*$ is the friction velocity defined as 
$v^*=\sqrt{{\tau _w}/{\rho}}$, where $\tau _w$ is the shear stress at the 
wall. For the same $Re$, we also calculate the Kolmogorov length scale as 
$\eta=({\nu^3}/{\epsilon})^{\frac{1}{4}}$, where $\epsilon$ is the
dissipation rate of the turbulent kinetic energy. We find that $y^+$ is less 
than $0.14$ throughout the surface, and the surface averaged value of the
ratio of the mesh element size with the Kolmogorov length scale is 
$3.15$, reflecting that the grid resolution near the surface is sufficient
to resolve the boundary layer and flow structures associated with its
transition. Further, we conducted a mesh convergence study 
where we compare the results from the original mesh with 
those obtained from meshes with higher grid resolution at 
several $Re$ across all the regimes.
Higher resolution meshes were generated by refining $N_\theta$ and $\Delta z$. 
We find that increasing the mesh resolution does not significantly affect the flow,
indicating that the mesh resolution considered here is 
appropriate for the $Re$ range. The convergence study of the mesh 
and its assessment to appropriately resolve the boundary 
layer transition is discussed in detail in our earlier 
work, \citep{chopra2021jfm}.

\section{Results}
\subsection{\textcolor{black}{The spatio-temporal dynamics of the shear layer instability}}
\begin{figure}
	\centerline
	{\includegraphics[width=1\textwidth]{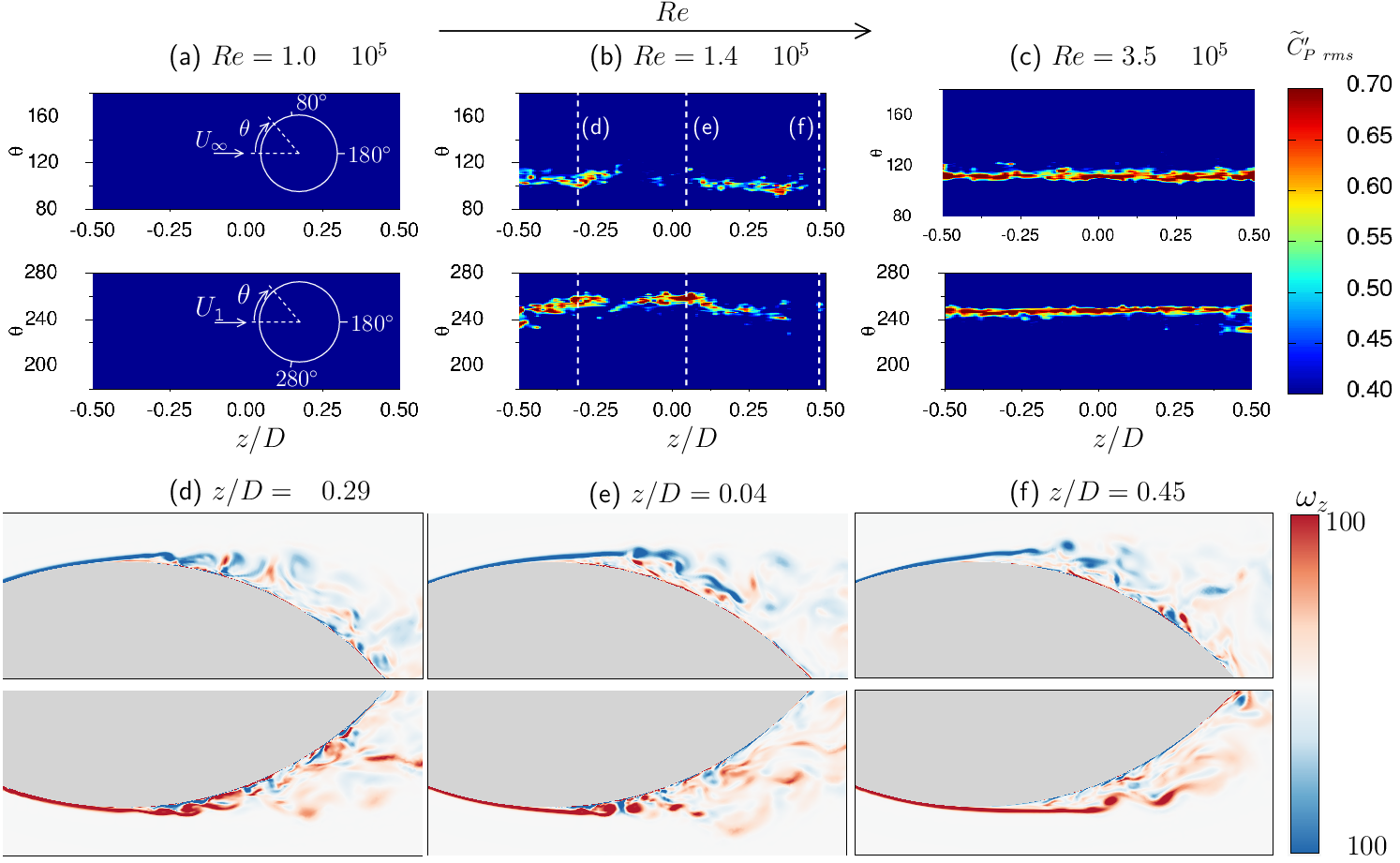}}
	\caption{Flow past a circular cylinder: 
    $z-\theta$ diagrams of instantaneous $\widetilde{C}'_{P~{rms}}(z,\theta,t)$ on top and bottom shoulders at $Re=$ 
    (a) $1.0\times10^5$ (subcritical), (b) $1.4\times10^5$ (critical), and
    (c) $3.5\times10^5$ (supercritical). The
	contours of the spanwise component of vorticity ($\omega_z$)
    for $Re=1.4\times10^5$ at stations marked in (b) 
    ${z}/{D}=$ (d) $-0.29$, (e) $0.04$, and (f) $0.45$.}
\label{fig:cprms_figure}
\end{figure}

\begin{figure}
	\centerline
.	{\includegraphics[width=1\textwidth]{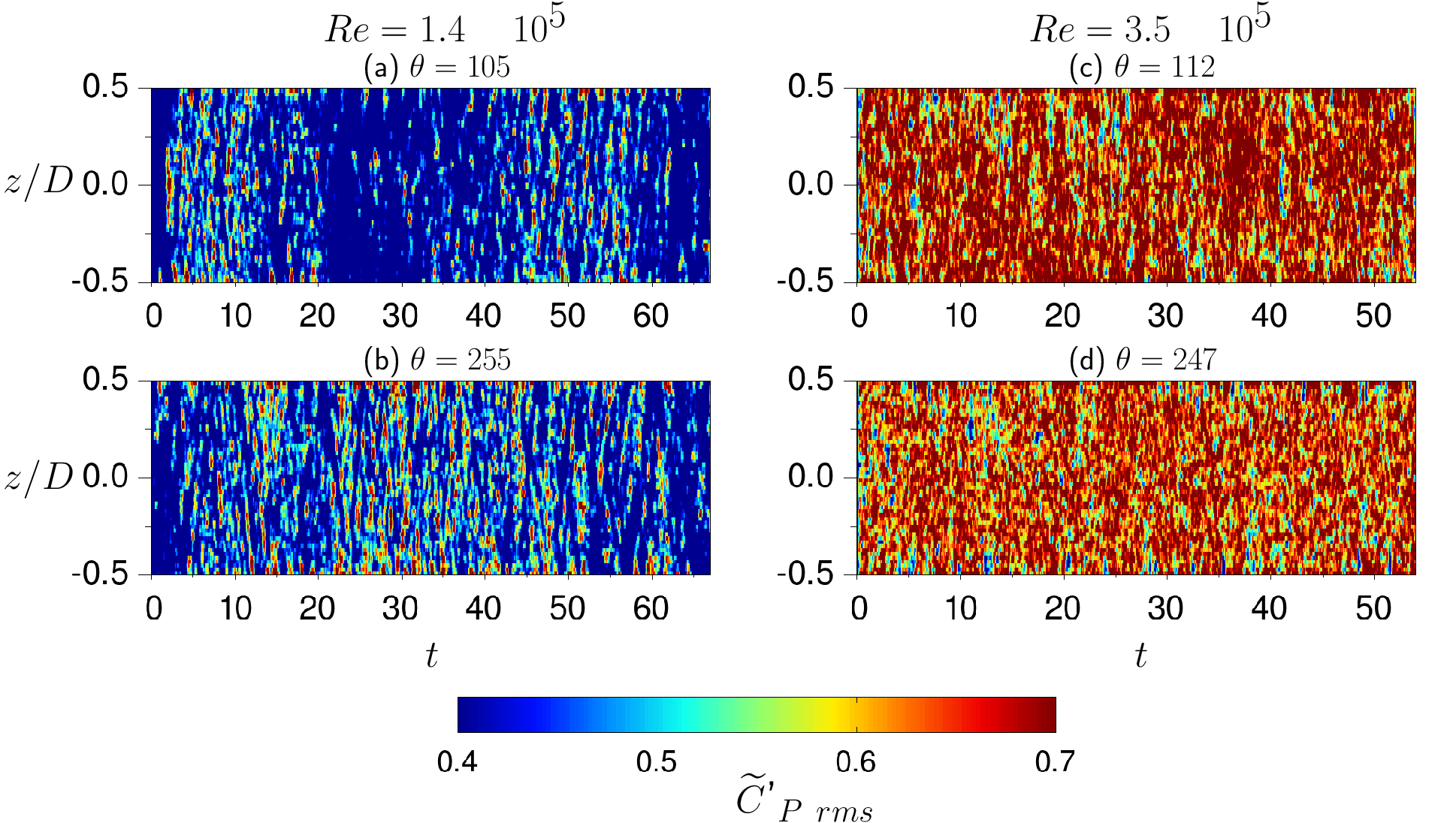}}
	\caption{\textcolor{black}{Flow past a circular cylinder: 
    $z-t$ diagrams of $\widetilde{C}'_{P~{rms}}(z,\theta,t)$ for $Re=1.4\times10^5$ at (a) $\theta=105^\circ$, (b) $\theta=255^\circ$, and
    $Re=3.5\times10^5$ at (c) $\theta=112^\circ$, (d) $\theta=247^\circ$.}}
\label{fig:cprms_figure_Re}
\end{figure}
\textcolor{black}
{
We study the evolution, with $Re$, of fluctuations in the coefficient of pressure on the surface of the cylinder due to vortices generated by the shear layer instability estimated using the double decomposition proposed by \citet{chopra2017intermittent}.
}
Figures \ref{fig:cprms_figure} (a-c) show the spanwise variation 
of $\widetilde{C}'_{P~{rms}}(\theta)$
on the upper and lower shoulders for instantaneous flow at time instants when 
the $C_L$ is the maximum during a vortex 
shedding cycle for $Re=1.0\times10^5$, $1.4\times10^5$ and $3.5\times10^5$ respectively.
Flow at $Re=1.0\times10^5$ is representative of flow in the subcritical regime. As expected, we do not 
observe high fluctuations at this $Re$ since the boundary layer 
separates in a laminar state without reattaching. 
$Re=1.4\times10^5$ being in the critical regime, we observe
regions of high $\widetilde{C}'_{P~{rms}}$ on both the upper and lower 
surfaces. The $\widetilde{C}'_{P~{rms}}$ is clustered and 
distributed incoherently along the span of the cylinder.
Also shown in figures \ref{fig:cprms_figure} (d)-(f) 
are the spanwise component of vorticity ($\omega_z$) fields at 
several span stations marked in (b).
At $z/D=-0.29$, $\widetilde{C}'_{P~{rms}}$ fluctuations are
high on both upper and lower shoulders, which is indicative of
turbulence generated due to shear layer instability.
We observe from the instantaneous $\omega_z$ field shown in figure 
\ref{fig:cprms_figure} (d) that, indeed, the shear layers 
separating from the upper and lower shoulders transition to a 
turbulent state and roll up into shear layer vortices close to 
the surface, causing the flow to reattach.
\citet{singh2005flow,cheng_pullin_samtaney_zhang_gao_2017}
and \citet{chopra2021jfm} showed that the velocity profile 
of the reattached flow in inner variables exhibit the 
signature of viscous sublayer close to the wall and that of 
the log-law for a certain region outwards. 
Further, \citet{cheng_pullin_samtaney_zhang_gao_2017} 
and \citet{chopra2021jfm} reported that the shape factor 
of the boundary layer drops rapidly to a low value after 
reattachment.
These characteristics confirm that the reattached flow
develops into a turbulent boundary layer. Therefore, we 
refer to such flow states as turbulent 
reattachment states \citep{chopra2021jfm}.

At $z=0.04D$, $\widetilde{C}'_{P~{rms}}$ fluctuations are 
low on the top shoulder.
We observe from the $\omega_z$ field at this span station 
(figure \ref{fig:cprms_figure} (e)) that the shear layer instability on the upper shoulder occurs 
relatively farther away from the surface, resulting in low 
fluctuations. Due to low turbulence generation near the surface, the flow does not reattach. Therefore, we refer to such flow states as no-reattachment states. On the other hand, fluctuations are high on the lower surface at $z=0.04D$. Correspondingly, the shear layer instability is close enough to 
cause flow reattachment in a turbulent state (see figure 
\ref{fig:cprms_figure} (e)). Both upper and lower shoulders are devoid of high $\widetilde{C}'_{P~{rms}}$ fluctuations at $z/D=0.45$. The fields of $\omega_z$ shown in figure \ref{fig:cprms_figure} (f) confirm that the flow on both shoulders is in a no-reattachment state.

\textcolor{black}{Figure \ref{fig:cprms_figure_Re} shows the span-time ($z-t$) diagrams of $\widetilde{C}'_{P~{rms}}$ at certain azimuthal locations on the upper and lower shoulders where the fluctuations due to shear layer instability are the strongest for $Re=1.4\times10^5$ and $3.5\times10^5$. This figure reveals the complex behavior of turbulent reattachment phenomena arising due to spatio-temporal intermittency of shear layer instability. At $Re=1.4\times10^5$, the $\widetilde{C}'_{P~{rms}}$ field is incoherent along the span and fragmented at all the time instants. Further, the fields of $\widetilde{C}'_{P~{rms}}$ fields appear to be random as we cannot identify any ordered regular patterns in the complex spatio-temporal dynamics.}

\textcolor{black}{
On the other hand, at $Re=3.5\times10^5$, which is in the supercritical regime, $\widetilde{C}'_{P~{rms}}$ is high throughout the field on both upper and lower shoulders (see figures \ref{fig:cprms_figure_Re}(c-d)). We observe from figure \ref{fig:cprms_figure} (c) that the fluctuations are relatively less clustered and fragmented, and flow undergoes turbulent reattachment all along the span. As a result, the turbulent reattachment is more coherent in the span-wise direction than at $Re=1.4\times10^5$ in the critical regime.
Next, we investigate the mechanism behind the transformation of the 
shear layer instability from a spanwise incoherent distribution to a 
coherent one using the complex network approach.}

\begin{figure}
	\centerline{\includegraphics[width=1.0\textwidth]{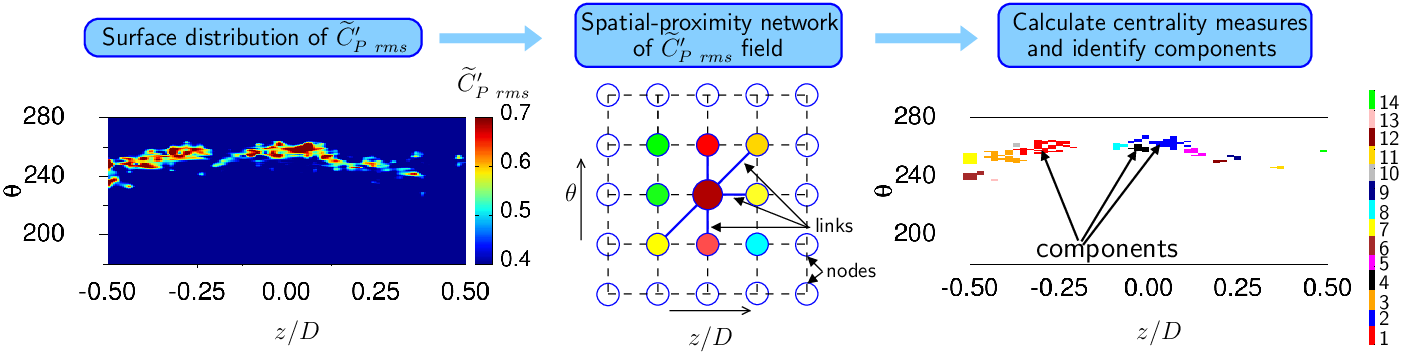}}
  \caption{Flow past a circular cylinder: an overview of the procedure adopted for the complex network analysis.}
\label{fig:network_schematic}
\end{figure}

\subsection{Complex network analysis of the flow field}
We study the evolution of clusters of boundary layer transition with $Re$ using a time-varying spatial-proximity network. 
We consider the vertices of the FEM mesh on the surface of the cylinder as the nodes of the network. \textcolor{black}{We treat strong  $\widetilde{C}'_{P~{rms}}$ fluctuations due to the shear layer instability as events and establish links when events are synchronized between two nodes.
As shown in figures \ref{fig:cprms_figure} (c) - (e), the shear layer instability occurs downstream of the shoulder of the cylinder.
Therefore, we restrict our region of interest from $80^{\circ}\leq\theta\leq180^{\circ}$ 
and $180^{\circ} \leq \theta \leq 280^{\circ}$ on the 
upper and lower surfaces. This region consists of $N=11373$ nodes on each surface.}

\subsubsection{Network construction methodology} \label{sec:NC_meth}
\par \textcolor{black}{ Figure \ref{fig:network_schematic} shows the procedure for constructing a time-varying spatial-proximity network. The link between two nodes $i$ and $j$ is established when the following two criteria are met: 
\begin{enumerate}[start=1,label={\bfseries C\arabic*.}]
\item \textit{$\widetilde{C}'_{P~{rms}}$ fluctuations at both nodes are higher than the threshold ($\tau$) simultaneously at time instant $t$: $\widetilde{C}'_{P~{rms}}(i,t) \geq  \tau \land \widetilde{C}'_{P~{rms}}(j,t) \geq \tau$}
\item \textit{\textcolor{black}{Both nodes are in spatial proximity: $\Delta z(i,j) \leq \tau_{\Delta z} \land \Delta \theta (i,j)\leq \tau_{\Delta \theta}$,  }}
\end{enumerate}
Where $\Delta z(i,j)$ is the spanwise spacing and $\Delta \theta (i,j)$ the azimuthal angle between nodes $i$ and $j$. $\tau_{\Delta z}$ and $\tau_{\Delta \theta}$ are thresholds in spanwise and circumferential directions. \textcolor{black}{Criterion $\mathbf{C1}$ incorporates the event description since it allows only those $\widetilde{C}'_{P~{rms}}$ fluctuations to be considered that are larger than the threshold $\tau$. It also enforces the condition that events should occur simultaneously in time.}
The $\widetilde{C}'_{P~{rms}}$ threshold is considered to be $\tau=0.6$. This value corresponds to relatively strong shear layer activity occurring close to the surface and leading to the flow reattachment in a turbulent state (see figure \ref{fig:cprms_figure}) across the entire range of $Re$ considered in this study. We investigate the sensitivity of $\tau$ on the network analysis and find that the effect of \textcolor{black}{$\tau$ is only quantitative and not qualitative}. The effect of $\tau$ is discussed in detail in Appendix \ref{app:cprms_tau}.}

The second schematic in figure \ref{fig:network_schematic} shows the network representation of the $\widetilde{C}'_{P~{rms}}$ field. The node under consideration is filled and enlarged, while the eligible nodes with which it can have a connection are filled, and their colors represent the intensity of $\widetilde{C}'_{P~{rms}}$.
The connections of the network are encoded in an adjacency matrix
($\mathbf{A}$) of size $N \times N$ whose elements: \textcolor{black}{
 	\begin{equation}
		A_{ij} = \begin{cases} \begin{array}{ll}
	    1,  & \mbox{if }\mathbf{C1} \mbox{ and } \mathbf{C2} \mbox{ are true}\\
	    0,  & \mbox{otherwise}\\
	\end{array}
	\end{cases}.
	\end{equation}
}

Network science provides an efficient framework for identifying clusters or islands in networks through the concept of connected components \citep{rokach2010data,krishnan2019emergence}. Connected components in a network are sub-groups of nodes wherein every node of a sub-group is connected to every other node via a path (sequence of links). On the other hand, nodes belonging to different sub-groups or components are not connected via a path \citep{barabasi2016network,newman2018networks}. We identify connected components in our time-varying spatial proximity networks. In the present context, they represent islands or clusters of high $\widetilde{C}'_{P~{rms}}$ where turbulent reattachment of flow occurs. \textcolor{black}{The spatial proximity constraint, i.e., the $\mathbf{C2}$ criterion, enables accurate prediction of the number of clusters in the flow field. We consider 
$\tau_{\Delta z}=0.02D$, and $\tau_{\Delta \theta}=0.45^\circ$. These values correspond to the length of the edges of the FEM mesh in the spanwise and circumferential direction on the surface of the cylinder.
Connected components identified in the spatial proximity network of $\widetilde{C}'_{P~{rms}}$ field are shown in the third column of figure \ref{fig:network_schematic}. We observe that clusters/islands of strong $\widetilde{C}'_{P~rms}$ fluctuations are identified as separated components.
We discuss the effect of relaxing the criterion $\mathbf{C2}$ by increasing $\tau_{\Delta z}$ and $\tau_{\Delta \theta}$ in Appendix \ref{app:span_tau}. Relaxation of this condition enables long-range connections, particularly between nodes belonging to different clusters of strong shear layer instability. As a result, two different clusters are counted as a single component, which is undesirable as it can lead to inaccurate estimation of cluster characteristics. Thus, we consider $\tau_{\Delta z}$ and $\tau_{\Delta \theta}$ corresponding to the spatial resolution in the finite element computations.}


 

\begin{figure}
	\centerline
	{\includegraphics[width=1\textwidth]{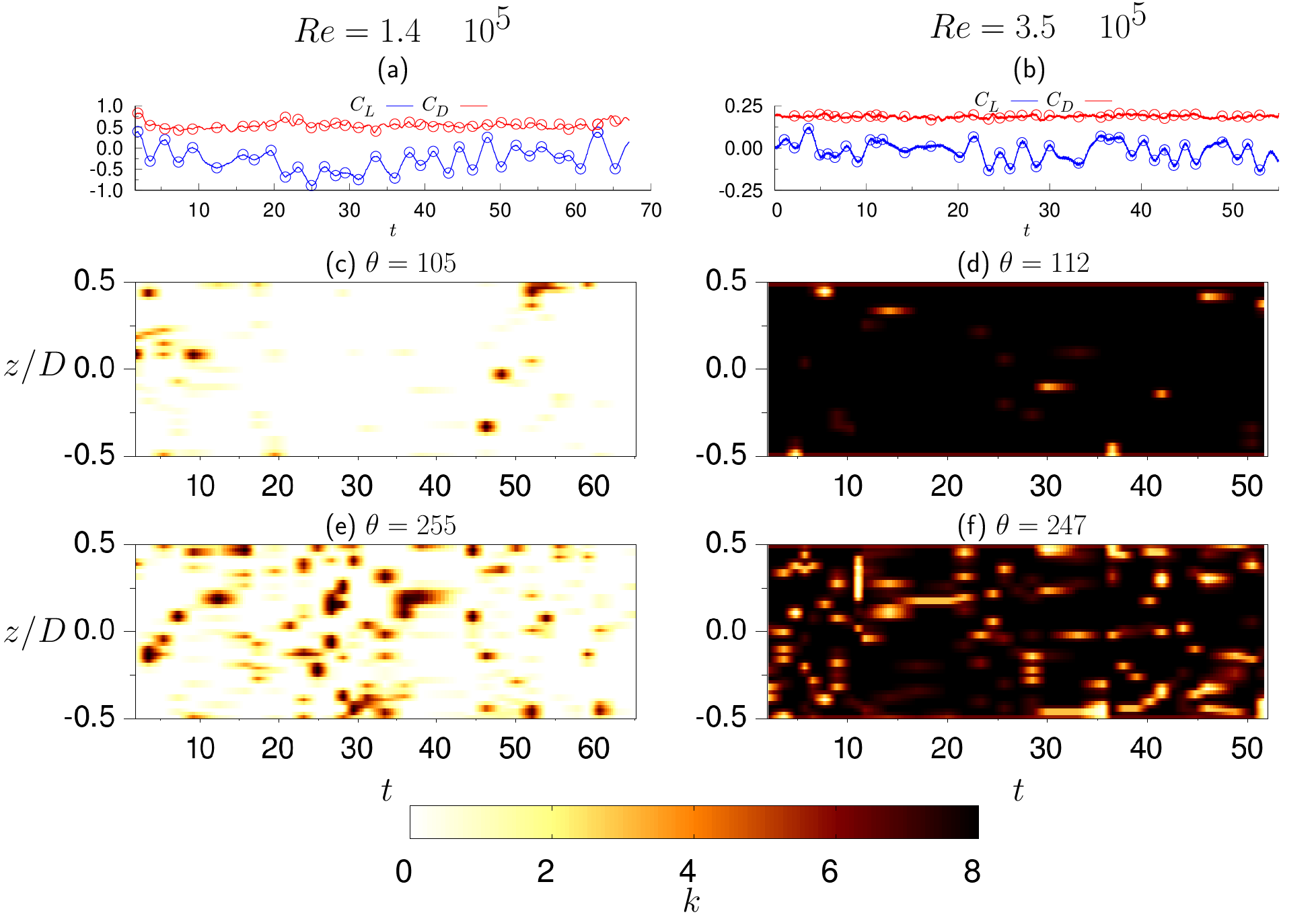}}
	\caption{\textcolor{black}{Flow past a circular cylinder: time-histories of coefficients of lift ($C_L$) and drag ($C_D$) for $Re=$ (a) $1.4\times10^5$ and (b) $3.5\times10^5$. The network analysis has been performed on time instants corresponding to maximum and minimum value $C_L$ in a vortex shedding cycle. These instants are marked with open circles in (a) and (b). Also shown are the $z-t$ diagrams of the degree ($k$ $(z,\theta,t)$) of spatial-proximity networks for $Re=1.4\times10^5$ at (c) $\theta=105^\circ$, (e) $\theta=255^\circ$, and $Re=3.5\times10^5$ at (d) $\theta=112^\circ$, (f) $\theta=247^\circ$ at time instants marked in (a) and (b) respectively.}}
\label{fig:degree_figure_Re}
\end{figure}

\subsubsection{Evolution of network measures with $Re$}    
    \textcolor{black}{We construct time-varying spatial-proximity networks at time instants corresponding to maximum and minimum values of the lift coefficient ($C_L$) in a vortex shedding cycle. 
    We estimate the number ($N_C$) of components, their average size ($S_{CA}$), the size of the largest component ($S_{CL}$), and the average degree ($D$) of spatial proximity networks at each time instant.
    Average degree is defined as $D=\frac{1}{N}\sum_{i=1}^{N}k_i$.
        Here, $k_i$ is the degree centrality, for node $i$, and is defined as $k_i=\sum_{j=1}^{N}A_{ij}$. It represents the number of direct connections or links of the node. In terms of the flow physics, it indicates the number of its neighbors with high fluctuations due to shear layer instability.}

    \textcolor{black}{Figure \ref{fig:degree_figure_Re} shows the $z-t$ diagrams of degree centrality ($k$) of networks constructed at time instants corresponding to a minimum and maximum value of $C_L$ in a vortex shedding cycle for $Re=1.4\times10^5$ and $3.5\times10^5$ at the same $\theta$ locations as in figure \ref{fig:cprms_figure_Re}. A high value of $k$ represents regions where  $\widetilde{C}'_{P~rms}$ fluctuations are strong. At $Re=1.4\times10^5$, regions of high $k$ appear intermittently and are fragmented, which implies that networks are sparse. On the other hand, at $Re=3.5\times10^5$, $k$ is consistently high, indicating that the networks are densely connected. We report time-averaged values of network measures.
    They are denoted with an over-bar ($\overline{\textcolor{white}{a}}$). 
    The time average is estimated for at least $18$ vortex shedding cycles.}
        

\begin{figure}
	\centerline{\includegraphics[width=1\textwidth]{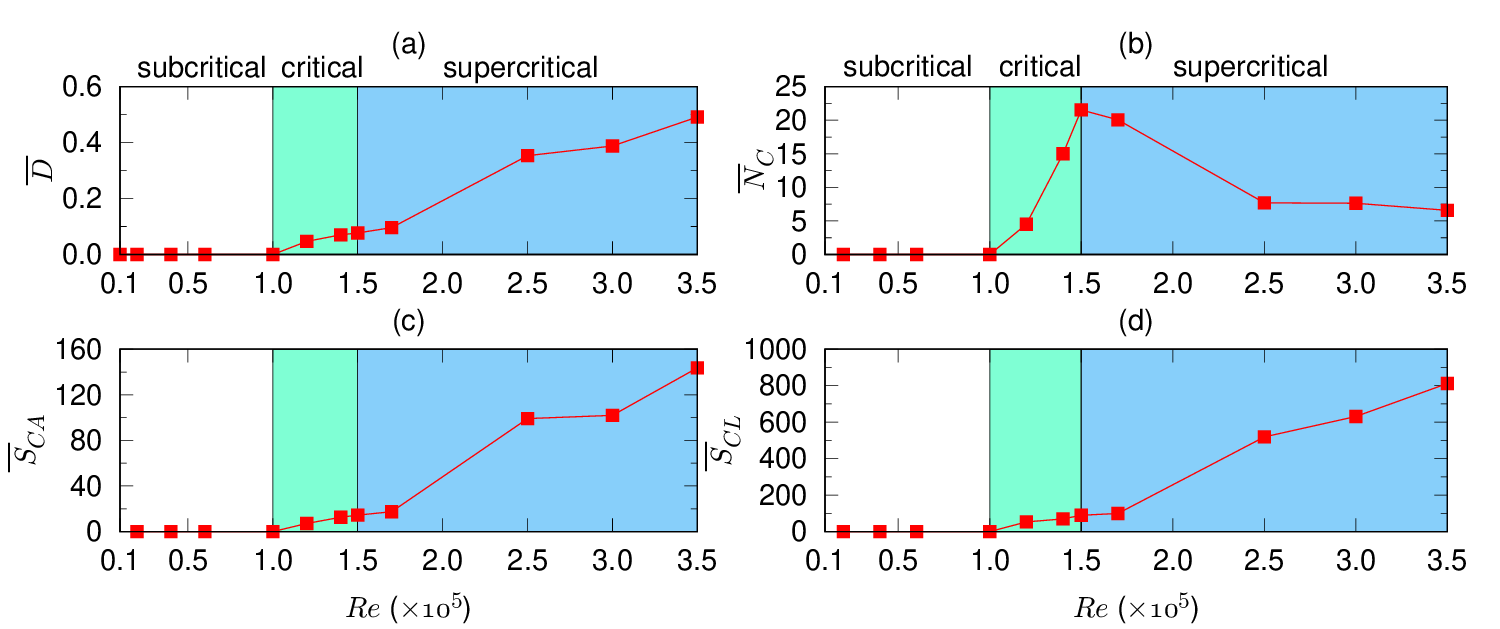}}
	\caption{Complex network analysis of flow past a circular cylinder: variation of time-averaged 
	(a) average degree ($\overline{D}$),
	(b) number of components ($\overline{N}_C$),
	(c) average size of components ($\overline{S}_{CA}$), and
	(d) size of the largest component ($\overline{S}_{CL}$) with $Re$.}
\label{fig:NC_SC}
\end{figure}

    
    Figures \ref{fig:NC_SC} (a-d) show the variation, with $Re$, of the $\overline{D}$, $\overline{N}_C$, $\overline{S}_{CA}$, and $\overline{S}_{CL}$,
    respectively. The average values from the lower and upper surfaces are reported here.
    As expected, all the quantities are zero in the subcritical regime \textcolor{black}{since $\widetilde{C}'_{P~rms}$ fluctuations on the surface due to the shear layer instability are negligible, and there is no flow reattachment in this regime.} $\overline{D}$ increases gradually with increase in $Re$
    in the critical and supercritical regimes. This indicates that regions of turbulent reattachment due to high shear layer instability expand with increase
    in $Re$. Variations of $\overline{N}_C$, $\overline{S}_{CA}$ and $\overline{S}_{CL}$ reveal interesting flow physics. These measures increase with $Re$ in the critical regime, indicating that the expansion of regions of turbulent reattachment happens through an increase in both the number and size of the clusters. In the supercritical regime, $\overline{N}_C$ decreases while $\overline{S}_{CA}$ and $\overline{S}_{CL}$ continue to increase with increase in $Re$. This reflects that while expanding, the turbulent reattachment clusters merge to form larger clusters. Consequently, their number decreases with increase in $Re$.

\begin{figure}
	\centerline{\includegraphics[width=1\textwidth]{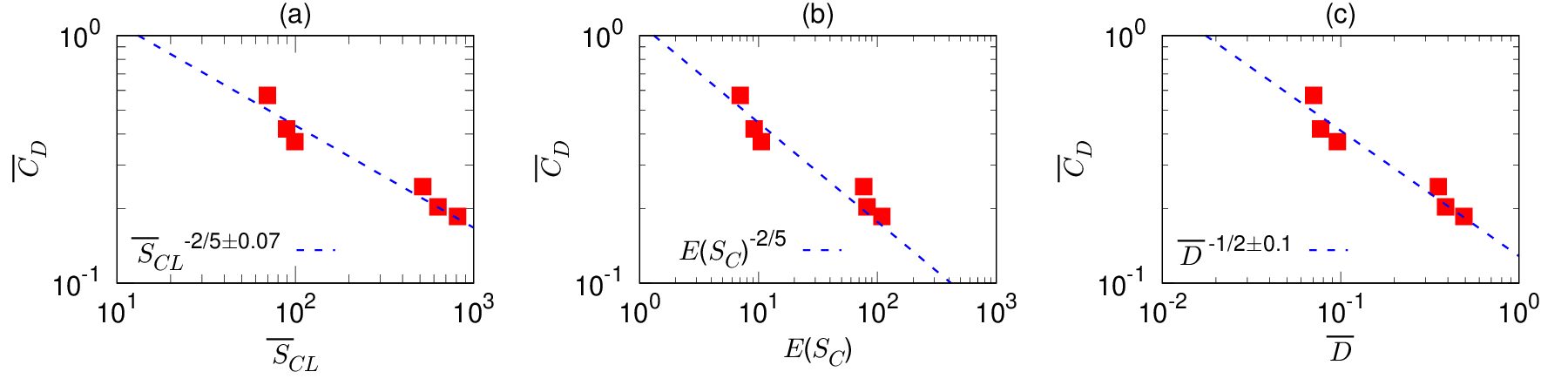}}
	\caption{\textcolor{black}{Complex network analysis of flow past a circular cylinder in the critical and supercritical regimes: variation of time-averaged drag coefficient ($\overline{C}_D$) with time-averaged (a) size of the largest component ($\overline{S}_{CL}$), (b) most probable component size ($E(S_C)$) and (c) average degree ($\overline{D}$). The best-fit curves are also shown via broken lines. \textcolor{black}{The $\overline{C}_D$ varies as ${\overline{S}_{CL}}^{-\frac{2}{5}}$ ($\chi^2=1.06\times10^{-2}$), $E(S_C)^{-\frac{2}{5}}$ and ${\overline{D}}^{-\frac{1}{2}}$ ($\chi^2=1.22\times10^{-2}$). Here, $\chi^2$ is the chi-squared goodness-of-fit measure estimated as the sum of squared differences between the actual data and that obtained from the power law distribution \citep{devore1995probability,williams2017gnuplot}. }}}
\label{fig:comp_size_CD}
\end{figure}

\subsubsection{Dependency of $\overline{C}_D$ on the characteristics of the clusters}
\textcolor{black}{The characteristics of clusters of strong fluctuations due to the shear layer instability, such as the time-averaged size of the largest cluster ($\overline{S}_{CL}$), and most probable cluster size $E(S_C)$ can be utilized to explain the variation of $\overline{C}_D$ in the critical and supercritical regime. Figures \ref{fig:comp_size_CD} (a) and (b) show the variation of $\overline{C}_D$ with $\overline{S}_{CL}$ and $E(S_C)$. The most probable component size is estimated as $E(S_C)=\sum S_C\times\mathbb{P}(S_C)$, where $\mathbb{P}(S_C)$ is the probability of the size of connected components ($S_C$). In probability theory, $E(S_C)$ is also known as the estimated value. $\mathbb{P}(S_C)$ is obtained from the probability distribution of $S_C$ estimated at all the time instants considered. Also shown, for reference, in figure \ref{fig:comp_size_CD} (c) is the variation of $\overline{C}_D$ with the $\overline{D}$ of the spatial-proximity networks.
We observe from figure \ref{fig:comp_size_CD} that, $\overline{C}_D$ decreases with increase in $\overline{S}_{CL}$, $E(S_C)$ and $\overline{D}$. As discussed in the preceding section, increase in $\overline{S}_{CL}$, $E(S_C)$ and $\overline{D}$ implies that the zone of strong fluctuations due to the shear layer instability on the cylinder surface is expanding, particularly in the spanwise direction. As a result, the zones of turbulent reattachment and delayed turbulent separation expand, leading to a decrease in $\overline{C}_D$ with increase in $Re$ \citep{achenbach1968distribution,son2011mechanism,chopra2021jfm}.
\textcolor{black}{We find that the variation of $\overline{C}_D$ with ${\overline{S}_{CL}}$, $E(S_C)$ and $\overline{D}$ follows power-law distribution. $\overline{C}_D$ varies with ${\overline{S}_{CL}}$ as $\overline{C}_D \propto {\overline{S}_{CL}}^{-\frac{2}{5}}$. This power law also fits the variation of $\overline{C}_D$ with $E(S_C)$ with good accuracy. Meanwhile, $\overline{C}_D$ varies with $\overline{D}$ as $\overline{C}_D \propto \overline{D}^{-\frac{1}{2}}$.}
}
\section{Conclusions}
The present study investigates the evolution of the spanwise characteristics of the flow with $Re$
in the critical and supercritical regimes via LES.
In these regimes, the separated shear layer transitions very close to the surface, leading to boundary layer 
transition. The enhanced turbulence enables the separated flow to reattach and develop 
into a turbulent boundary layer. \textcolor{black}{Using the surface pressure fluctuations due to the shear layer instability estimated via double decomposition proposed by \citet{chopra2017intermittent} we explore the complex spatio-temporal dynamics of the turbulent reattachment of the boundary layer. } We observe that turbulent reattachment does not occur at all span locations simultaneously in the early critical regime. Meanwhile, at higher $Re$ in the supercritical regime, 
the turbulent reattachment is no longer clustered as it occurs coherently along the span of the cylinder.

\textcolor{black}{We study the evolution of the spatio-temporal dynamics of the flow with $Re$ using a novel complex networks approach that incorporates strong fluctuations in the surface pressure generated by vortices due to the shear layer instability as extreme events. We construct a time-varying spatial proximity network where the nodes are the discrete vertices of the FEM mesh on the surface of the cylinder. Links representing interaction between two nodes are established when the nodes are in spatial proximity to each other and experience an extreme event, corresponding to pressure fluctuations larger than a specified threshold, simultaneously in time.}
\par \textcolor{black}{This analysis enables estimating the number and size of clusters of turbulent reattachment in the flow field and studying their evolution with $Re$.}
We observe that the connected components in this network represent clusters of turbulent reattachment in the flow field.
The characteristics of the connected components bring out the underlying mechanism of
transformation of the spatial distribution of turbulent reattachment from a clustered to a coherent one with increase in $Re$. \textcolor{black}{It is found that the number and size of components govern the variation of the time-averaged coefficient of drag ($\overline{C}_D$) in the critical and supercritical regimes.}
Initially, in the early critical regime, the components are sparse, and their size is small. 
With increase in $Re$, they grow both in size and number. This is
indicative of the appearance of bigger clusters of 
turbulent reattachment at more spanwise locations with increase in $Re$.
In the supercritical regime, clusters of turbulent reattachment merge to form bigger clusters, resulting in an improvement in the spanwise coherence of the flow. Merging of clusters also results in a drop in the number of clusters. Expansion of the region of turbulent reattachment leads to a decrease in the $\overline{C}_D$ with increase in $Re$. \textcolor{black}{$\overline{C}_D$ exhibits power law variation with the most probable cluster size ($\overline{C}_D \propto E(S_C)^{-\frac{2}{5}}$) and the largest cluster size ($\overline{C}_D \propto {\overline{S}_{CL}}^{-\frac{2}{5}}$) across the wide range of $Re$ in the critical and supercritical regimes, indicating the importance of cluster characteristics in these regimes.}

\textcolor{black}{This study presents a novel approach based on complex network theory to study turbulent flows from a new perspective. Through the proposed approach, one can incorporate the event description \citep{narsimha1995turbulence} of turbulent fluctuations rather than spectral description where fluctuations are decomposed as waves. The spatial proximity approach proposed here can be extended to study extreme events in different flow properties in other turbulent flows. However, this approach does not consider non-linear interactions between extreme events prevalent in turbulent flows \citep{iacobello2021large}. Here, links are established if events occur simultaneously, which can be considered a type of linear interaction. Non-linearity can be incorporated if statistical similarity measures such as event synchronization \citep{quiroga2002event} are utilized for quantifying interactions. Such analysis will give deeper insights into the dynamics of turbulent flows and address several unresolved research questions raised by \citet{narsimha1995turbulence}. For example, how well are the extreme events correlated across the spatio-temporal flow field? And how are extreme events distributed in terms of magnitude, duration, and arrival times? These interesting questions can be explored in a future study.} 


%

\begin{acknowledgments}
The authors acknowledge the use of HPC, PARAM Sanganak
facilities at the Indian Institute of Technology Kanpur and Cray
XC-40, Shaheen, at King Abdullah University of Science and
Technology, Saudi Arabia. G.C. and R.I.S. are grateful for the funding provided by the Office of 
Naval Research Global (O.N.R.G) under grant No. SP21221684AEONRG002696 (contract monitor: Dr. D. M. Tepaske).
\end{acknowledgments}
\section*{Declaration of interests}
The authors report no conflict of interest.
\section*{Data Availability Statement}
The data supporting this study's findings 
are available from the corresponding author upon reasonable request.

\appendix
\section{Effect of the threshold of $\widetilde{C}'_{P~{rms}}$ ($\tau$)} \label{app:cprms_tau}
\begin{figure}
	\centerline{\includegraphics[width=1\textwidth]{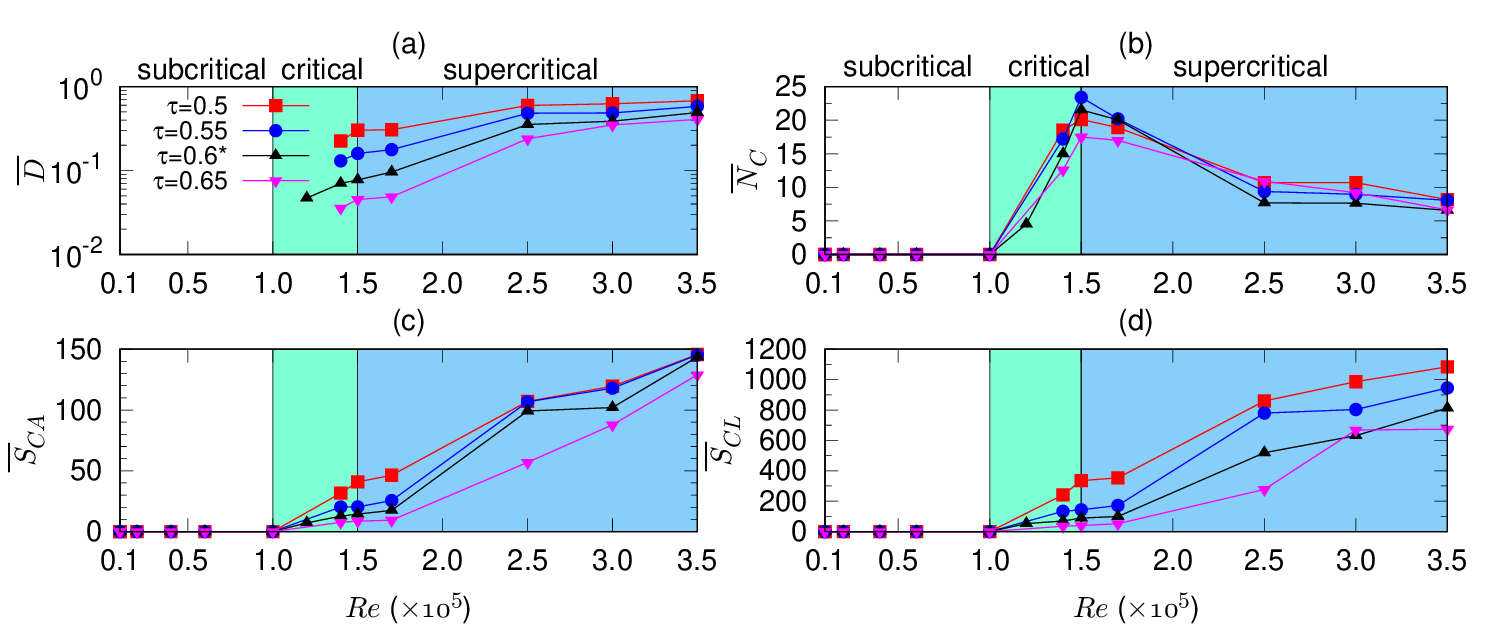}}
	\caption{\textcolor{black}{Complex network analysis of flow past a circular cylinder: variation of time-averaged 
	(a) average degree ($\overline{D}$),
	(b) number of components ($\overline{N}_C$),
	(c) average size of components ($\overline{S}_{CA}$), and
	(d) size of the largest component ($\overline{S}_{CL}$) with $Re$ for spatial-proximity networks constructed using several values of threshold ($\tau$) for $\widetilde{C}'_{P~{rms}}(z,\theta,t)$.}}
\label{fig:net_comp_thresh}
\end{figure}

\textcolor{black}{We study the effect of $\widetilde{C}'_{P~{rms}}$ threshold ($\tau$) on our analysis by comparing network measures for spatial-proximity networks constructed using $\tau=0.5,~0.55,~0.6$ and $0.65$ while keeping $\tau_{\Delta z}$ and $\tau_{\Delta \theta}$ constant at $0.02D$ and $0.45^\circ$ respectively.} Figures \ref{fig:net_comp_thresh}(a-d) show the variation of time-averaged network measures, average degree ($\overline{D}$), number of components ($\overline{N}_C$), the average size of components ($\overline{S}_{CA}$), and size of the largest component ($\overline{S}_{CL}$) respectively, with $Re$ for several values of $\tau$. We observe from figure \ref{fig:net_comp_thresh} that trends of network measures with the $Re$ for different $\tau$ are similar. Thus, the effect of varying $\tau$ is only quantitative and not qualitative. The conclusions drawn from the analysis presented in this paper are consistent and robust with respect to variation in $\tau$. In the present study, we consider $\tau=0.6$ since this value is associated with large fluctuations due to shear layer instability causing turbulent reattachment across the entire $Re$ range considered (see figure \ref{fig:cprms_figure}).

\section{Effect of $\tau_{\Delta z}$ and $\tau_{\Delta \theta}$} \label{app:span_tau}
\begin{figure}
	\centerline{\includegraphics[width=1\textwidth]{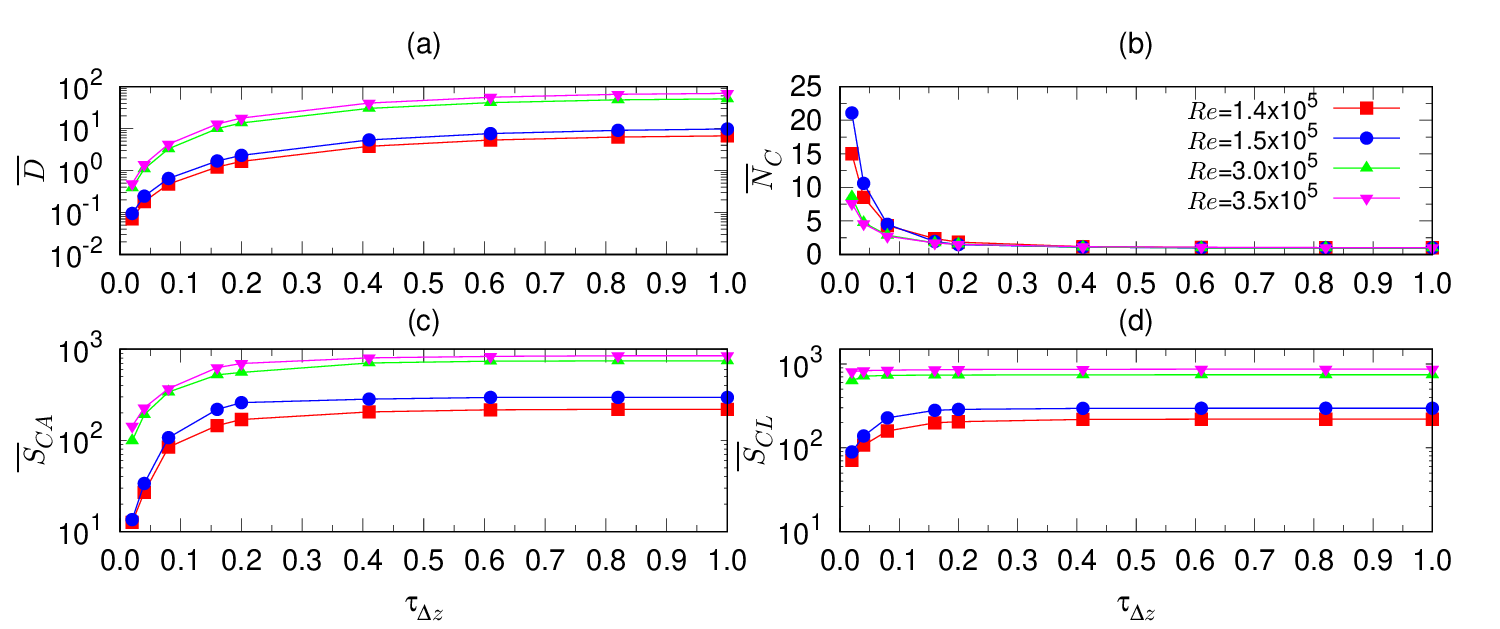}}
	\caption{\textcolor{black}{Complex network analysis of flow past a circular cylinder in critical and supercritical regimes: variation of time-averaged 
	(a) average degree ($\overline{D}$),
	(b) number of components ($\overline{N}_C$),
	(c) average size of components ($\overline{S}_{CA}$), and
	(d) size of the largest component ($\overline{S}_{CL}$) with the threshold for link distance in the spanwise direction ($\tau_{\Delta z}$) for constructing spatial-proximity networks at several $Re$.}}
\label{fig:net_comp_span_thresh}
\end{figure}

\textcolor{black}{In the present study, we impose the spatial proximity constraint (the criterion $\mathbf{C2}$) for constructing networks. According to the criterion $\mathbf{C2}$, two nodes can be connected only if they are in spatial proximity. 
    We select an appropriate proximity condition by investigating the effect of varying $\tau_{\Delta z}$ and $\tau_{\Delta \theta}$ on network measures and component characteristics. 
    Figures \ref{fig:net_comp_span_thresh} 
    (a-d) show the variation of $\overline{D}$, $\overline{N}_C$, $\overline{S}_{CA}$ and $\overline{S}_{CL}$ with $\tau_{\Delta z}$ for several $Re$ in the critical and supercritical regimes. We observe from figure \ref{fig:net_comp_span_thresh} (a) that $\overline{D}$ increases with increase in $\tau_{\Delta z}$. Such a variation is expected because relaxing the spatial proximity condition allows non-local and long-range links, making the network denser.} 

\begin{figure}
	\centerline{\includegraphics[width=1.1\textwidth]{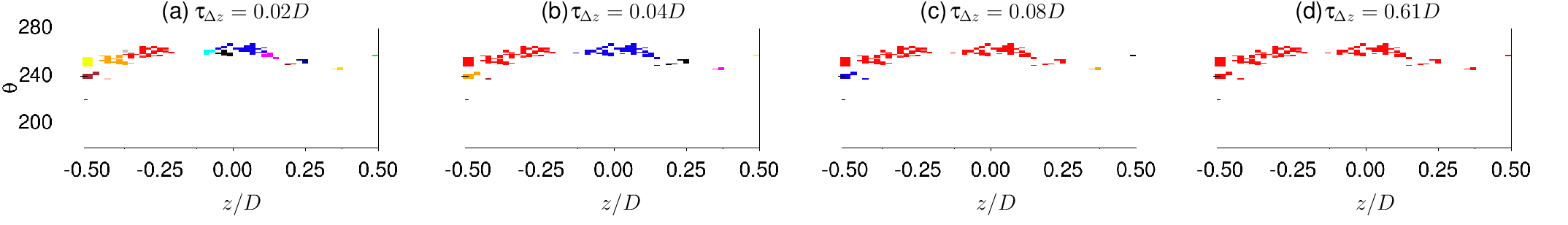}}
	\caption{\textcolor{black}{Complex network analysis of $Re=1.4\times10^5$ flow past a circular cylinder: connected components in spatial-proximity networks of  $\widetilde{C}'_{P~rms}$ field shown in figure \ref{fig:network_schematic} constructed using $\tau_{\Delta z}=$ (a) $0.02D$, (b) $0.04D$, (d) $0.08D$ and (d) $0.61D$. Each component is color-coded for clarity. }}
\label{fig:comp_span}
\end{figure}

\textcolor{black}{
Relaxing spatial proximity constraints also affects the characteristics of the components in the network. Their number decreases (see figure \ref{fig:net_comp_span_thresh} (b)) and size increases (see figures \ref{fig:net_comp_span_thresh} (c) and (d)) with increase in $\tau_{\Delta z}$.
The underlying phenomenon can be understood from figure \ref{fig:comp_span} which shows the connected components in networks constructed for $\widetilde{C}'_{P~rms}$ field shown in figure \ref{fig:network_schematic} using $\tau_{\Delta z}=0.02D,~0.04D$, $0.08D$ and $0.61D$. Increasing $\tau_{\Delta z}$ allows the formation of long-range links in the network, especially between nodes that are physically in separate clusters. This scenario results in separate clusters getting counted as one single component. In fact, beyond $\tau_{\Delta z}\approx 0.61$, the network becomes fully connected, resulting in only one single large-sized component ($\overline{N}_C=1$, see figures \ref{fig:net_comp_span_thresh} and \ref{fig:comp_span}) which is undesirable and inconsistent with the actual flow physics. As discussed earlier, turbulent reattachment due to shear layer instability occurs in clusters, especially in the critical regime. Further, our aim is to identify these clusters via components in the network and study their evolution. Therefore, we consider $\tau_{\Delta z}=0.02D$ and $\tau_{\Delta \theta}=0.45^\circ$ for performing network analysis since these values result in the largest number of components and most accurate network representation of actual flow physics.}

\bibliography{thesis}

\begin{thebibliography}{46}%
\makeatletter
\providecommand \@ifxundefined [1]{%
 \@ifx{#1\undefined}
}%
\providecommand \@ifnum [1]{%
 \ifnum #1\expandafter \@firstoftwo
 \else \expandafter \@secondoftwo
 \fi
}%
\providecommand \@ifx [1]{%
 \ifx #1\expandafter \@firstoftwo
 \else \expandafter \@secondoftwo
 \fi
}%
\providecommand \natexlab [1]{#1}%
\providecommand \enquote  [1]{``#1''}%
\providecommand \bibnamefont  [1]{#1}%
\providecommand \bibfnamefont [1]{#1}%
\providecommand \citenamefont [1]{#1}%
\providecommand \href@noop [0]{\@secondoftwo}%
\providecommand \href [0]{\begingroup \@sanitize@url \@href}%
\providecommand \@href[1]{\@@startlink{#1}\@@href}%
\providecommand \@@href[1]{\endgroup#1\@@endlink}%
\providecommand \@sanitize@url [0]{\catcode `\\12\catcode `\$12\catcode
  `\&12\catcode `\#12\catcode `\^12\catcode `\_12\catcode `\%12\relax}%
\providecommand \@@startlink[1]{}%
\providecommand \@@endlink[0]{}%
\providecommand \url  [0]{\begingroup\@sanitize@url \@url }%
\providecommand \@url [1]{\endgroup\@href {#1}{\urlprefix }}%
\providecommand \urlprefix  [0]{URL }%
\providecommand \Eprint [0]{\href }%
\providecommand \doibase [0]{http://dx.doi.org/}%
\providecommand \selectlanguage [0]{\@gobble}%
\providecommand \bibinfo  [0]{\@secondoftwo}%
\providecommand \bibfield  [0]{\@secondoftwo}%
\providecommand \translation [1]{[#1]}%
\providecommand \BibitemOpen [0]{}%
\providecommand \bibitemStop [0]{}%
\providecommand \bibitemNoStop [0]{.\EOS\space}%
\providecommand \EOS [0]{\spacefactor3000\relax}%
\providecommand \BibitemShut  [1]{\csname bibitem#1\endcsname}%
\let\auto@bib@innerbib\@empty
\bibitem [{\citenamefont {Achenbach}(1968)}]{achenbach1968distribution}%
  \BibitemOpen
  \bibfield  {author} {\bibinfo {author} {\bibnamefont {Achenbach},
  \bibfnamefont {E.}},\ }\bibfield  {title} {\enquote {\bibinfo {title}
  {Distribution of local pressure and skin friction around a circular cylinder
  in cross-flow up to {R}e= 5$\times$10$^6$},}\ }\href@noop {} {\bibfield
  {journal} {\bibinfo  {journal} {J. Fluid Mech.}\ }\textbf {\bibinfo {volume}
  {34}},\ \bibinfo {pages} {625--639} (\bibinfo {year} {1968})}\BibitemShut
  {NoStop}%
\bibitem [{\citenamefont {Arakeri}(2021)}]{arakeri2021turbulence}%
  \BibitemOpen
  \bibfield  {author} {\bibinfo {author} {\bibnamefont {Arakeri}, \bibfnamefont
  {J.~H.}},\ }\bibfield  {title} {\enquote {\bibinfo {title} {Turbulence: Waves
  or events?}}\ }\href@noop {} {\bibfield  {journal} {\bibinfo  {journal}
  {Resonance}\ }\textbf {\bibinfo {volume} {24}},\ \bibinfo {pages}
  {1465--1471} (\bibinfo {year} {2021})}\BibitemShut {NoStop}%
\bibitem [{\citenamefont {Barab{\'a}si}(2016)}]{barabasi2016network}%
  \BibitemOpen
  \bibfield  {author} {\bibinfo {author} {\bibnamefont {Barab{\'a}si},
  \bibfnamefont {A.~L.}},\ }\href@noop {} {\emph {\bibinfo {title} {Network
  Science}}}\ (\bibinfo  {publisher} {Cambridge University Press},\ \bibinfo
  {year} {2016})\BibitemShut {NoStop}%
\bibitem [{\citenamefont {Cadot}\ \emph {et~al.}(2015)\citenamefont {Cadot},
  \citenamefont {Desai}, \citenamefont {Mittal}, \citenamefont {Saxena},\ and\
  \citenamefont {Chandra}}]{cadot2015statistics}%
  \BibitemOpen
  \bibfield  {author} {\bibinfo {author} {\bibnamefont {Cadot}, \bibfnamefont
  {O.}}, \bibinfo {author} {\bibnamefont {Desai}, \bibfnamefont {A.}}, \bibinfo
  {author} {\bibnamefont {Mittal}, \bibfnamefont {S.}}, \bibinfo {author}
  {\bibnamefont {Saxena}, \bibfnamefont {S.}}, \ and\ \bibinfo {author}
  {\bibnamefont {Chandra}, \bibfnamefont {B.}},\ }\bibfield  {title} {\enquote
  {\bibinfo {title} {Statistics and dynamics of the boundary layer
  reattachments during the drag crisis transitions of a circular cylinder},}\
  }\href@noop {} {\bibfield  {journal} {\bibinfo  {journal} {Phys. Fluids}\
  }\textbf {\bibinfo {volume} {27}},\ \bibinfo {pages} {014101} (\bibinfo
  {year} {2015})}\BibitemShut {NoStop}%
\bibitem [{\citenamefont {Cheng}\ \emph {et~al.}(2017)\citenamefont {Cheng},
  \citenamefont {Pullin}, \citenamefont {Samtaney}, \citenamefont {Zhang},\
  and\ \citenamefont {Gao}}]{cheng_pullin_samtaney_zhang_gao_2017}%
  \BibitemOpen
  \bibfield  {author} {\bibinfo {author} {\bibnamefont {Cheng}, \bibfnamefont
  {W.}}, \bibinfo {author} {\bibnamefont {Pullin}, \bibfnamefont {D.~I.}},
  \bibinfo {author} {\bibnamefont {Samtaney}, \bibfnamefont {R.}}, \bibinfo
  {author} {\bibnamefont {Zhang}, \bibfnamefont {W.}}, \ and\ \bibinfo {author}
  {\bibnamefont {Gao}, \bibfnamefont {W.}},\ }\bibfield  {title} {\enquote
  {\bibinfo {title} {Large-eddy simulation of flow over a cylinder with
  ${R}e_{D}$ from $3.9\times 10^{3}$ to $8.5\times 10^{5}$: a skin-friction
  perspective},}\ }\href@noop {} {\bibfield  {journal} {\bibinfo  {journal} {J.
  Fluid Mech.}\ }\textbf {\bibinfo {volume} {820}},\ \bibinfo {pages}
  {121–158} (\bibinfo {year} {2017})}\BibitemShut {NoStop}%
\bibitem [{\citenamefont {Chopra}\ and\ \citenamefont
  {Mittal}(2017)}]{chopra2017intermittent}%
  \BibitemOpen
  \bibfield  {author} {\bibinfo {author} {\bibnamefont {Chopra}, \bibfnamefont
  {G.}}\ and\ \bibinfo {author} {\bibnamefont {Mittal}, \bibfnamefont {S.}},\
  }\bibfield  {title} {\enquote {\bibinfo {title} {The intermittent nature of
  the laminar separation bubble on a cylinder in uniform flow},}\ }\href@noop
  {} {\bibfield  {journal} {\bibinfo  {journal} {Comput. \& Fluids}\ }\textbf
  {\bibinfo {volume} {142}},\ \bibinfo {pages} {118--127} (\bibinfo {year}
  {2017})}\BibitemShut {NoStop}%
\bibitem [{\citenamefont {Chopra}\ and\ \citenamefont
  {Mittal}(2022{\natexlab{a}})}]{chopra2022effect}%
  \BibitemOpen
  \bibfield  {author} {\bibinfo {author} {\bibnamefont {Chopra}, \bibfnamefont
  {G.}}\ and\ \bibinfo {author} {\bibnamefont {Mittal}, \bibfnamefont {S.}},\
  }\bibfield  {title} {\enquote {\bibinfo {title} {The effect of trip wire on
  transition of boundary layer on a cylinder},}\ }\href@noop {} {\bibfield
  {journal} {\bibinfo  {journal} {Phys. Fluids}\ }\textbf {\bibinfo {volume}
  {34}},\ \bibinfo {pages} {054103} (\bibinfo {year}
  {2022}{\natexlab{a}})}\BibitemShut {NoStop}%
\bibitem [{\citenamefont {Chopra}\ and\ \citenamefont
  {Mittal}(2022{\natexlab{b}})}]{chopra2021jfm}%
  \BibitemOpen
  \bibfield  {author} {\bibinfo {author} {\bibnamefont {Chopra}, \bibfnamefont
  {G.}}\ and\ \bibinfo {author} {\bibnamefont {Mittal}, \bibfnamefont {S.}},\
  }\bibfield  {title} {\enquote {\bibinfo {title} {Secondary vortex, laminar
  separation bubble and vortex shedding in flow past a low aspect ratio
  circular cylinder},}\ }\href@noop {} {\bibfield  {journal} {\bibinfo
  {journal} {J. Fluid Mech.}\ }\textbf {\bibinfo {volume} {930}},\ \bibinfo
  {pages} {A12} (\bibinfo {year} {2022}{\natexlab{b}})}\BibitemShut {NoStop}%
\bibitem [{\citenamefont {Chopra}\ and\ \citenamefont
  {Mittal}(2023)}]{chopra2023laminar}%
  \BibitemOpen
  \bibfield  {author} {\bibinfo {author} {\bibnamefont {Chopra}, \bibfnamefont
  {G.}}\ and\ \bibinfo {author} {\bibnamefont {Mittal}, \bibfnamefont {S.}},\
  }\bibfield  {title} {\enquote {\bibinfo {title} {Laminar separation bubble on
  a rotating cylinder in uniform flow},}\ }\href@noop {} {\bibfield  {journal}
  {\bibinfo  {journal} {Phys. Fluids}\ }\textbf {\bibinfo {volume} {35}},\
  \bibinfo {pages} {044105} (\bibinfo {year} {2023})}\BibitemShut {NoStop}%
\bibitem [{\citenamefont {Desai}\ and\ \citenamefont
  {Mittal}(2022)}]{desai2022effect}%
  \BibitemOpen
  \bibfield  {author} {\bibinfo {author} {\bibnamefont {Desai}, \bibfnamefont
  {A.}}\ and\ \bibinfo {author} {\bibnamefont {Mittal}, \bibfnamefont {S.}},\
  }\bibfield  {title} {\enquote {\bibinfo {title} {Effect of free stream
  turbulence on the topology of laminar separation bubble on a sphere},}\
  }\href@noop {} {\bibfield  {journal} {\bibinfo  {journal} {J. Fluid Mech.}\
  }\textbf {\bibinfo {volume} {948}},\ \bibinfo {pages} {A28} (\bibinfo {year}
  {2022})}\BibitemShut {NoStop}%
\bibitem [{\citenamefont {Desai}, \citenamefont {Mittal},\ and\ \citenamefont
  {Mittal}(2020)}]{desai2020vortexshedding}%
  \BibitemOpen
  \bibfield  {author} {\bibinfo {author} {\bibnamefont {Desai}, \bibfnamefont
  {A.}}, \bibinfo {author} {\bibnamefont {Mittal}, \bibfnamefont {S.}}, \ and\
  \bibinfo {author} {\bibnamefont {Mittal}, \bibfnamefont {S.}},\ }\bibfield
  {title} {\enquote {\bibinfo {title} {Experimental investigation of vortex
  shedding past circular cylinder in the high subcritical regime},}\
  }\href@noop {} {\bibfield  {journal} {\bibinfo  {journal} {Phys. Fluids}\
  }\textbf {\bibinfo {volume} {32}},\ \bibinfo {pages} {014105} (\bibinfo
  {year} {2020})}\BibitemShut {NoStop}%
\bibitem [{\citenamefont {Deshpande}\ \emph {et~al.}(2017)\citenamefont
  {Deshpande}, \citenamefont {Kanti}, \citenamefont {Desai},\ and\
  \citenamefont {Mittal}}]{deshpande2017intermittency}%
  \BibitemOpen
  \bibfield  {author} {\bibinfo {author} {\bibnamefont {Deshpande},
  \bibfnamefont {R.}}, \bibinfo {author} {\bibnamefont {Kanti}, \bibfnamefont
  {V.}}, \bibinfo {author} {\bibnamefont {Desai}, \bibfnamefont {A.}}, \ and\
  \bibinfo {author} {\bibnamefont {Mittal}, \bibfnamefont {S.}},\ }\bibfield
  {title} {\enquote {\bibinfo {title} {Intermittency of laminar separation
  bubble on a sphere during drag crisis},}\ }\href@noop {} {\bibfield
  {journal} {\bibinfo  {journal} {J. Fluid Mech.}\ }\textbf {\bibinfo {volume}
  {812}},\ \bibinfo {pages} {815--840} (\bibinfo {year} {2017})}\BibitemShut
  {NoStop}%
\bibitem [{\citenamefont {Devore}(1995)}]{devore1995probability}%
  \BibitemOpen
  \bibfield  {author} {\bibinfo {author} {\bibnamefont {Devore}, \bibfnamefont
  {J.~L.}},\ }\href@noop {} {\emph {\bibinfo {title} {Probability and
  Statistics for Engineering and the Sciences}}},\ Vol.~\bibinfo {volume} {5}\
  (\bibinfo  {publisher} {Duxbury Press Belmont},\ \bibinfo {year}
  {1995})\BibitemShut {NoStop}%
\bibitem [{\citenamefont {Fernex}, \citenamefont {Noack},\ and\ \citenamefont
  {Semaan}(2021)}]{doi:10.1126/sciadv.abf5006}%
  \BibitemOpen
  \bibfield  {author} {\bibinfo {author} {\bibnamefont {Fernex}, \bibfnamefont
  {D.}}, \bibinfo {author} {\bibnamefont {Noack}, \bibfnamefont {B.~R.}}, \
  and\ \bibinfo {author} {\bibnamefont {Semaan}, \bibfnamefont {R.}},\
  }\bibfield  {title} {\enquote {\bibinfo {title} {Cluster-based network
  modeling—from snapshots to complex dynamical systems},}\ }\href {\doibase
  10.1126/sciadv.abf5006} {\bibfield  {journal} {\bibinfo  {journal} {Science
  Advances}\ }\textbf {\bibinfo {volume} {7}},\ \bibinfo {pages} {eabf5006}
  (\bibinfo {year} {2021})},\ \Eprint
  {http://arxiv.org/abs/https://www.science.org/doi/pdf/10.1126/sciadv.abf5006}
  {https://www.science.org/doi/pdf/10.1126/sciadv.abf5006} \BibitemShut
  {NoStop}%
\bibitem [{\citenamefont {Forouzi~Feshalami}\ \emph {et~al.}(2022)\citenamefont
  {Forouzi~Feshalami}, \citenamefont {He}, \citenamefont {Scarano},
  \citenamefont {Gan},\ and\ \citenamefont {Morton}}]{forouzi2022review}%
  \BibitemOpen
  \bibfield  {author} {\bibinfo {author} {\bibnamefont {Forouzi~Feshalami},
  \bibfnamefont {B.}}, \bibinfo {author} {\bibnamefont {He}, \bibfnamefont
  {S.}}, \bibinfo {author} {\bibnamefont {Scarano}, \bibfnamefont {F.}},
  \bibinfo {author} {\bibnamefont {Gan}, \bibfnamefont {L.}}, \ and\ \bibinfo
  {author} {\bibnamefont {Morton}, \bibfnamefont {C.}},\ }\bibfield  {title}
  {\enquote {\bibinfo {title} {A review of experiments on stationary bluff body
  wakes},}\ }\href@noop {} {\bibfield  {journal} {\bibinfo  {journal} {Physics
  of Fluids}\ }\textbf {\bibinfo {volume} {34}} (\bibinfo {year}
  {2022})}\BibitemShut {NoStop}%
\bibitem [{\citenamefont {Hadjighasem}\ \emph {et~al.}(2016)\citenamefont
  {Hadjighasem}, \citenamefont {Karrasch}, \citenamefont {Teramoto},\ and\
  \citenamefont {Haller}}]{PhysRevE.93.063107}%
  \BibitemOpen
  \bibfield  {author} {\bibinfo {author} {\bibnamefont {Hadjighasem},
  \bibfnamefont {A.}}, \bibinfo {author} {\bibnamefont {Karrasch},
  \bibfnamefont {D.}}, \bibinfo {author} {\bibnamefont {Teramoto},
  \bibfnamefont {H.}}, \ and\ \bibinfo {author} {\bibnamefont {Haller},
  \bibfnamefont {G.}},\ }\bibfield  {title} {\enquote {\bibinfo {title}
  {Spectral-clustering approach to lagrangian vortex detection},}\ }\href
  {\doibase 10.1103/PhysRevE.93.063107} {\bibfield  {journal} {\bibinfo
  {journal} {Phys. Rev. E}\ }\textbf {\bibinfo {volume} {93}},\ \bibinfo
  {pages} {063107} (\bibinfo {year} {2016})}\BibitemShut {NoStop}%
\bibitem [{\citenamefont {Iacobello}, \citenamefont {Kaiser},\ and\
  \citenamefont {Rival}(2022)}]{iacobello2022load}%
  \BibitemOpen
  \bibfield  {author} {\bibinfo {author} {\bibnamefont {Iacobello},
  \bibfnamefont {G.}}, \bibinfo {author} {\bibnamefont {Kaiser}, \bibfnamefont
  {F.}}, \ and\ \bibinfo {author} {\bibnamefont {Rival}, \bibfnamefont
  {D.~E.}},\ }\bibfield  {title} {\enquote {\bibinfo {title} {Load estimation
  in unsteady flows from sparse pressure measurements: Application of
  transition networks to experimental data},}\ }\href@noop {} {\bibfield
  {journal} {\bibinfo  {journal} {Physics of Fluids}\ }\textbf {\bibinfo
  {volume} {34}} (\bibinfo {year} {2022})}\BibitemShut {NoStop}%
\bibitem [{\citenamefont {Iacobello}, \citenamefont {Ridolfi},\ and\
  \citenamefont {Scarsoglio}(2021{\natexlab{a}})}]{iacobello2021large}%
  \BibitemOpen
  \bibfield  {author} {\bibinfo {author} {\bibnamefont {Iacobello},
  \bibfnamefont {G.}}, \bibinfo {author} {\bibnamefont {Ridolfi}, \bibfnamefont
  {L.}}, \ and\ \bibinfo {author} {\bibnamefont {Scarsoglio}, \bibfnamefont
  {S.}},\ }\bibfield  {title} {\enquote {\bibinfo {title} {Large-to-small scale
  frequency modulation analysis in wall-bounded turbulence via visibility
  networks},}\ }\href@noop {} {\bibfield  {journal} {\bibinfo  {journal} {J.
  Fluid Mech.}\ }\textbf {\bibinfo {volume} {918}},\ \bibinfo {pages} {A13}
  (\bibinfo {year} {2021}{\natexlab{a}})}\BibitemShut {NoStop}%
\bibitem [{\citenamefont {Iacobello}, \citenamefont {Ridolfi},\ and\
  \citenamefont {Scarsoglio}(2021{\natexlab{b}})}]{iacobello2021review}%
  \BibitemOpen
  \bibfield  {author} {\bibinfo {author} {\bibnamefont {Iacobello},
  \bibfnamefont {G.}}, \bibinfo {author} {\bibnamefont {Ridolfi}, \bibfnamefont
  {L.}}, \ and\ \bibinfo {author} {\bibnamefont {Scarsoglio}, \bibfnamefont
  {S.}},\ }\bibfield  {title} {\enquote {\bibinfo {title} {A review on
  turbulent and vortical flow analyses via complex networks},}\ }\href@noop {}
  {\bibfield  {journal} {\bibinfo  {journal} {Physica A}\ }\textbf {\bibinfo
  {volume} {563}},\ \bibinfo {pages} {125476} (\bibinfo {year}
  {2021}{\natexlab{b}})}\BibitemShut {NoStop}%
\bibitem [{\citenamefont {Iacobello}\ \emph {et~al.}(2019)\citenamefont
  {Iacobello}, \citenamefont {Scarsoglio}, \citenamefont {Kuerten},\ and\
  \citenamefont {Ridolfi}}]{iacobello_scarsoglio_kuerten_ridolfi_2019}%
  \BibitemOpen
  \bibfield  {author} {\bibinfo {author} {\bibnamefont {Iacobello},
  \bibfnamefont {G.}}, \bibinfo {author} {\bibnamefont {Scarsoglio},
  \bibfnamefont {S.}}, \bibinfo {author} {\bibnamefont {Kuerten}, \bibfnamefont
  {J.~G.~M.}}, \ and\ \bibinfo {author} {\bibnamefont {Ridolfi}, \bibfnamefont
  {L.}},\ }\bibfield  {title} {\enquote {\bibinfo {title} {Lagrangian network
  analysis of turbulent mixing},}\ }\href {\doibase 10.1017/jfm.2019.79}
  {\bibfield  {journal} {\bibinfo  {journal} {Journal of Fluid Mechanics}\
  }\textbf {\bibinfo {volume} {865}},\ \bibinfo {pages} {546–562} (\bibinfo
  {year} {2019})}\BibitemShut {NoStop}%
\bibitem [{\citenamefont {Kim}\ \emph {et~al.}(2014)\citenamefont {Kim},
  \citenamefont {Choi}, \citenamefont {Park},\ and\ \citenamefont
  {Yoo}}]{kim_choi_park_yoo_2014}%
  \BibitemOpen
  \bibfield  {author} {\bibinfo {author} {\bibnamefont {Kim}, \bibfnamefont
  {J.}}, \bibinfo {author} {\bibnamefont {Choi}, \bibfnamefont {H.}}, \bibinfo
  {author} {\bibnamefont {Park}, \bibfnamefont {H.}}, \ and\ \bibinfo {author}
  {\bibnamefont {Yoo}, \bibfnamefont {J.~Y.}},\ }\bibfield  {title} {\enquote
  {\bibinfo {title} {Inverse {M}agnus effect on a rotating sphere: when and
  why},}\ }\href {\doibase 10.1017/jfm.2014.428} {\bibfield  {journal}
  {\bibinfo  {journal} {J. Fluid Mech.}\ }\textbf {\bibinfo {volume} {754}},\
  \bibinfo {pages} {R2} (\bibinfo {year} {2014})}\BibitemShut {NoStop}%
\bibitem [{\citenamefont {Krishnan}\ \emph {et~al.}(2019)\citenamefont
  {Krishnan}, \citenamefont {Sujith}, \citenamefont {Marwan},\ and\
  \citenamefont {Kurths}}]{krishnan2019emergence}%
  \BibitemOpen
  \bibfield  {author} {\bibinfo {author} {\bibnamefont {Krishnan},
  \bibfnamefont {A.}}, \bibinfo {author} {\bibnamefont {Sujith}, \bibfnamefont
  {R.~I.}}, \bibinfo {author} {\bibnamefont {Marwan}, \bibfnamefont {N.}}, \
  and\ \bibinfo {author} {\bibnamefont {Kurths}, \bibfnamefont {J.}},\
  }\bibfield  {title} {\enquote {\bibinfo {title} {On the emergence of large
  clusters of acoustic power sources at the onset of thermoacoustic instability
  in a turbulent combustor},}\ }\href@noop {} {\bibfield  {journal} {\bibinfo
  {journal} {J. Fluid Mech.}\ }\textbf {\bibinfo {volume} {874}},\ \bibinfo
  {pages} {455--482} (\bibinfo {year} {2019})}\BibitemShut {NoStop}%
\bibitem [{\citenamefont {Krishnan}\ \emph {et~al.}(2021)\citenamefont
  {Krishnan}, \citenamefont {Sujith}, \citenamefont {Marwan},\ and\
  \citenamefont {Kurths}}]{krishnan2021suppression}%
  \BibitemOpen
  \bibfield  {author} {\bibinfo {author} {\bibnamefont {Krishnan},
  \bibfnamefont {A.}}, \bibinfo {author} {\bibnamefont {Sujith}, \bibfnamefont
  {R.~I.}}, \bibinfo {author} {\bibnamefont {Marwan}, \bibfnamefont {N.}}, \
  and\ \bibinfo {author} {\bibnamefont {Kurths}, \bibfnamefont {J.}},\
  }\bibfield  {title} {\enquote {\bibinfo {title} {Suppression of
  thermoacoustic instability by targeting the hubs of the turbulent networks in
  a bluff body stabilized combustor},}\ }\href@noop {} {\bibfield  {journal}
  {\bibinfo  {journal} {J. Fluid Mech.}\ }\textbf {\bibinfo {volume} {916}},\
  \bibinfo {pages} {A20} (\bibinfo {year} {2021})}\BibitemShut {NoStop}%
\bibitem [{\citenamefont {Landau}\ and\ \citenamefont
  {Lifshitz}(1982)}]{landau1982fluid}%
  \BibitemOpen
  \bibfield  {author} {\bibinfo {author} {\bibnamefont {Landau}, \bibfnamefont
  {L.}}\ and\ \bibinfo {author} {\bibnamefont {Lifshitz}, \bibfnamefont {E.}},\
  }\href@noop {} {\emph {\bibinfo {title} {Fluid Mechanics}}}\ (\bibinfo
  {publisher} {Pergaman, New York, USA},\ \bibinfo {year} {1982})\BibitemShut
  {NoStop}%
\bibitem [{\citenamefont {Macau}(2018)}]{macau2018mathematical}%
  \BibitemOpen
  \bibfield  {author} {\bibinfo {author} {\bibnamefont {Macau}, \bibfnamefont
  {E.~E.}},\ }\href@noop {} {\emph {\bibinfo {title} {A mathematical modeling
  approach from nonlinear dynamics to complex systems}}},\ Vol.~\bibinfo
  {volume} {22}\ (\bibinfo  {publisher} {Springer},\ \bibinfo {year}
  {2018})\BibitemShut {NoStop}%
\bibitem [{\citenamefont {Meena}, \citenamefont {Nair},\ and\ \citenamefont
  {Taira}(2018)}]{meena2018network}%
  \BibitemOpen
  \bibfield  {author} {\bibinfo {author} {\bibnamefont {Meena}, \bibfnamefont
  {M.~G.}}, \bibinfo {author} {\bibnamefont {Nair}, \bibfnamefont {A.~G.}}, \
  and\ \bibinfo {author} {\bibnamefont {Taira}, \bibfnamefont {K.}},\
  }\bibfield  {title} {\enquote {\bibinfo {title} {Network community-based
  model reduction for vortical flows},}\ }\href@noop {} {\bibfield  {journal}
  {\bibinfo  {journal} {Phys. Rev. E}\ }\textbf {\bibinfo {volume} {97}},\
  \bibinfo {pages} {063103} (\bibinfo {year} {2018})}\BibitemShut {NoStop}%
\bibitem [{\citenamefont {Meena}\ and\ \citenamefont
  {Taira}(2021)}]{meena2021identifying}%
  \BibitemOpen
  \bibfield  {author} {\bibinfo {author} {\bibnamefont {Meena}, \bibfnamefont
  {M.~G.}}\ and\ \bibinfo {author} {\bibnamefont {Taira}, \bibfnamefont {K.}},\
  }\bibfield  {title} {\enquote {\bibinfo {title} {Identifying vortical network
  connectors for turbulent flow modification},}\ }\href@noop {} {\bibfield
  {journal} {\bibinfo  {journal} {J. Fluid Mech.}\ }\textbf {\bibinfo {volume}
  {915}},\ \bibinfo {pages} {A10} (\bibinfo {year} {2021})}\BibitemShut
  {NoStop}%
\bibitem [{\citenamefont {Narasimha}(1995)}]{narsimha1995turbulence}%
  \BibitemOpen
  \bibfield  {author} {\bibinfo {author} {\bibnamefont {Narasimha},
  \bibfnamefont {R.}},\ }\bibfield  {title} {\enquote {\bibinfo {title}
  {Turbulence: waves or events?}}\ }\href
  {http://www.jstor.org/stable/24096168} {\bibfield  {journal} {\bibinfo
  {journal} {Current Science}\ }\textbf {\bibinfo {volume} {68}},\ \bibinfo
  {pages} {33--38} (\bibinfo {year} {1995})}\BibitemShut {NoStop}%
\bibitem [{\citenamefont {Narasimha}\ and\ \citenamefont
  {Kailas}(1990)}]{narasimha1990turbulent}%
  \BibitemOpen
  \bibfield  {author} {\bibinfo {author} {\bibnamefont {Narasimha},
  \bibfnamefont {R.}}\ and\ \bibinfo {author} {\bibnamefont {Kailas},
  \bibfnamefont {S.}},\ }\bibfield  {title} {\enquote {\bibinfo {title}
  {Turbulent bursts in the atmosphere},}\ }\href@noop {} {\bibfield  {journal}
  {\bibinfo  {journal} {Atmospheric Environment. Part A. General Topics}\
  }\textbf {\bibinfo {volume} {24}},\ \bibinfo {pages} {1635--1645} (\bibinfo
  {year} {1990})}\BibitemShut {NoStop}%
\bibitem [{\citenamefont {Newman}(2018)}]{newman2018networks}%
  \BibitemOpen
  \bibfield  {author} {\bibinfo {author} {\bibnamefont {Newman}, \bibfnamefont
  {M.}},\ }\href@noop {} {\emph {\bibinfo {title} {Networks}}}\ (\bibinfo
  {publisher} {Oxford University press},\ \bibinfo {year} {2018})\BibitemShut
  {NoStop}%
\bibitem [{\citenamefont {Nicoud}\ \emph {et~al.}(2011)\citenamefont {Nicoud},
  \citenamefont {Toda}, \citenamefont {Cabrit}, \citenamefont {Bose},\ and\
  \citenamefont {Lee}}]{nicoud2011using}%
  \BibitemOpen
  \bibfield  {author} {\bibinfo {author} {\bibnamefont {Nicoud}, \bibfnamefont
  {F.}}, \bibinfo {author} {\bibnamefont {Toda}, \bibfnamefont {H.~B.}},
  \bibinfo {author} {\bibnamefont {Cabrit}, \bibfnamefont {O.}}, \bibinfo
  {author} {\bibnamefont {Bose}, \bibfnamefont {S.}}, \ and\ \bibinfo {author}
  {\bibnamefont {Lee}, \bibfnamefont {J.}},\ }\bibfield  {title} {\enquote
  {\bibinfo {title} {Using singular values to build a subgrid-scale model for
  large eddy simulations},}\ }\href@noop {} {\bibfield  {journal} {\bibinfo
  {journal} {Phys. Fluids}\ }\textbf {\bibinfo {volume} {23}},\ \bibinfo
  {pages} {085106} (\bibinfo {year} {2011})}\BibitemShut {NoStop}%
\bibitem [{\citenamefont {Padberg-Gehle}\ and\ \citenamefont
  {Schneide}(2017)}]{npg-24-661-2017}%
  \BibitemOpen
  \bibfield  {author} {\bibinfo {author} {\bibnamefont {Padberg-Gehle},
  \bibfnamefont {K.}}\ and\ \bibinfo {author} {\bibnamefont {Schneide},
  \bibfnamefont {C.}},\ }\bibfield  {title} {\enquote {\bibinfo {title}
  {Network-based study of lagrangian transport and mixing},}\ }\href {\doibase
  10.5194/npg-24-661-2017} {\bibfield  {journal} {\bibinfo  {journal}
  {Nonlinear Processes in Geophysics}\ }\textbf {\bibinfo {volume} {24}},\
  \bibinfo {pages} {661--671} (\bibinfo {year} {2017})}\BibitemShut {NoStop}%
\bibitem [{\citenamefont {Prasad}\ and\ \citenamefont
  {Williamson}(1997)}]{prasad1997instability}%
  \BibitemOpen
  \bibfield  {author} {\bibinfo {author} {\bibnamefont {Prasad}, \bibfnamefont
  {A.}}\ and\ \bibinfo {author} {\bibnamefont {Williamson}, \bibfnamefont
  {C.}},\ }\bibfield  {title} {\enquote {\bibinfo {title} {The instability of
  the shear layer separating from a bluff body},}\ }\href@noop {} {\bibfield
  {journal} {\bibinfo  {journal} {J. Fluid Mech.}\ }\textbf {\bibinfo {volume}
  {333}},\ \bibinfo {pages} {375--402} (\bibinfo {year} {1997})}\BibitemShut
  {NoStop}%
\bibitem [{\citenamefont {Quiroga}, \citenamefont {Kreuz},\ and\ \citenamefont
  {Grassberger}(2002)}]{quiroga2002event}%
  \BibitemOpen
  \bibfield  {author} {\bibinfo {author} {\bibnamefont {Quiroga}, \bibfnamefont
  {R.~Q.}}, \bibinfo {author} {\bibnamefont {Kreuz}, \bibfnamefont {T.}}, \
  and\ \bibinfo {author} {\bibnamefont {Grassberger}, \bibfnamefont {P.}},\
  }\bibfield  {title} {\enquote {\bibinfo {title} {Event synchronization: a
  simple and fast method to measure synchronicity and time delay patterns},}\
  }\href@noop {} {\bibfield  {journal} {\bibinfo  {journal} {Physical Review
  E}\ }\textbf {\bibinfo {volume} {66}},\ \bibinfo {pages} {041904} (\bibinfo
  {year} {2002})}\BibitemShut {NoStop}%
\bibitem [{\citenamefont {Rokach}\ and\ \citenamefont
  {Maimon}(2010)}]{rokach2010data}%
  \BibitemOpen
  \bibfield  {author} {\bibinfo {author} {\bibnamefont {Rokach}, \bibfnamefont
  {L.}}\ and\ \bibinfo {author} {\bibnamefont {Maimon}, \bibfnamefont {O.}},\
  }\href@noop {} {\emph {\bibinfo {title} {Data mining and knowledge discovery
  handbook}}}\ (\bibinfo  {publisher} {Springer New York},\ \bibinfo {year}
  {2010})\BibitemShut {NoStop}%
\bibitem [{\citenamefont {Roshko}(1961)}]{roshko1961experiments}%
  \BibitemOpen
  \bibfield  {author} {\bibinfo {author} {\bibnamefont {Roshko}, \bibfnamefont
  {A.}},\ }\bibfield  {title} {\enquote {\bibinfo {title} {Experiments on the
  flow past a circular cylinder at very high {R}eynolds number},}\ }\href@noop
  {} {\bibfield  {journal} {\bibinfo  {journal} {J. Fluid Mech.}\ }\textbf
  {\bibinfo {volume} {10}},\ \bibinfo {pages} {345--356} (\bibinfo {year}
  {1961})}\BibitemShut {NoStop}%
\bibitem [{\citenamefont {Scarsoglio}, \citenamefont {Iacobello},\ and\
  \citenamefont {Ridolfi}(2016)}]{scarsoglio2016complex}%
  \BibitemOpen
  \bibfield  {author} {\bibinfo {author} {\bibnamefont {Scarsoglio},
  \bibfnamefont {S.}}, \bibinfo {author} {\bibnamefont {Iacobello},
  \bibfnamefont {G.}}, \ and\ \bibinfo {author} {\bibnamefont {Ridolfi},
  \bibfnamefont {L.}},\ }\bibfield  {title} {\enquote {\bibinfo {title}
  {Complex networks unveiling spatial patterns in turbulence},}\ }\href@noop {}
  {\bibfield  {journal} {\bibinfo  {journal} {Intl J. Bifurcation and Chaos}\
  }\textbf {\bibinfo {volume} {26}},\ \bibinfo {pages} {1650223} (\bibinfo
  {year} {2016})}\BibitemShut {NoStop}%
\bibitem [{\citenamefont {Shri~Vignesh}\ \emph {et~al.}(2022)\citenamefont
  {Shri~Vignesh}, \citenamefont {Tandon}, \citenamefont {Kasthuri},\ and\
  \citenamefont {Sujith}}]{shri2022complex}%
  \BibitemOpen
  \bibfield  {author} {\bibinfo {author} {\bibnamefont {Shri~Vignesh},
  \bibfnamefont {K.}}, \bibinfo {author} {\bibnamefont {Tandon}, \bibfnamefont
  {S.}}, \bibinfo {author} {\bibnamefont {Kasthuri}, \bibfnamefont {P.}}, \
  and\ \bibinfo {author} {\bibnamefont {Sujith}, \bibfnamefont {R.}},\
  }\bibfield  {title} {\enquote {\bibinfo {title} {A complex network framework
  for studying particle-laden flows},}\ }\href@noop {} {\bibfield  {journal}
  {\bibinfo  {journal} {Physics of Fluids}\ }\textbf {\bibinfo {volume} {34}}
  (\bibinfo {year} {2022})}\BibitemShut {NoStop}%
\bibitem [{\citenamefont {Singh}\ and\ \citenamefont
  {Mittal}(2005)}]{singh2005flow}%
  \BibitemOpen
  \bibfield  {author} {\bibinfo {author} {\bibnamefont {Singh}, \bibfnamefont
  {S.}}\ and\ \bibinfo {author} {\bibnamefont {Mittal}, \bibfnamefont {S.}},\
  }\bibfield  {title} {\enquote {\bibinfo {title} {Flow past a cylinder: shear
  layer instability and drag crisis},}\ }\href@noop {} {\bibfield  {journal}
  {\bibinfo  {journal} {Intl J. Numer. Meth. Fluids}\ }\textbf {\bibinfo
  {volume} {47}},\ \bibinfo {pages} {75--98} (\bibinfo {year}
  {2005})}\BibitemShut {NoStop}%
\bibitem [{\citenamefont {Son}\ \emph {et~al.}(2011)\citenamefont {Son},
  \citenamefont {Choi}, \citenamefont {Jeon},\ and\ \citenamefont
  {Choi}}]{son2011mechanism}%
  \BibitemOpen
  \bibfield  {author} {\bibinfo {author} {\bibnamefont {Son}, \bibfnamefont
  {K.}}, \bibinfo {author} {\bibnamefont {Choi}, \bibfnamefont {J.}}, \bibinfo
  {author} {\bibnamefont {Jeon}, \bibfnamefont {W.-P.}}, \ and\ \bibinfo
  {author} {\bibnamefont {Choi}, \bibfnamefont {H.}},\ }\bibfield  {title}
  {\enquote {\bibinfo {title} {Mechanism of drag reduction by a surface trip
  wire on a sphere},}\ }\href@noop {} {\bibfield  {journal} {\bibinfo
  {journal} {J. Fluid Mech.}\ }\textbf {\bibinfo {volume} {672}},\ \bibinfo
  {pages} {411} (\bibinfo {year} {2011})}\BibitemShut {NoStop}%
\bibitem [{\citenamefont {Taira}\ and\ \citenamefont
  {Nair}(2022)}]{taira2022network}%
  \BibitemOpen
  \bibfield  {author} {\bibinfo {author} {\bibnamefont {Taira}, \bibfnamefont
  {K.}}\ and\ \bibinfo {author} {\bibnamefont {Nair}, \bibfnamefont {A.~G.}},\
  }\bibfield  {title} {\enquote {\bibinfo {title} {Network-based analysis of
  fluid flows: Progress and outlook},}\ }\href@noop {} {\bibfield  {journal}
  {\bibinfo  {journal} {Prog. Aerosp. Sci.}\ }\textbf {\bibinfo {volume}
  {131}},\ \bibinfo {pages} {100823} (\bibinfo {year} {2022})}\BibitemShut
  {NoStop}%
\bibitem [{\citenamefont {Taira}, \citenamefont {Nair},\ and\ \citenamefont
  {Brunton}(2016)}]{taira2016network}%
  \BibitemOpen
  \bibfield  {author} {\bibinfo {author} {\bibnamefont {Taira}, \bibfnamefont
  {K.}}, \bibinfo {author} {\bibnamefont {Nair}, \bibfnamefont {A.~G.}}, \ and\
  \bibinfo {author} {\bibnamefont {Brunton}, \bibfnamefont {S.~L.}},\
  }\bibfield  {title} {\enquote {\bibinfo {title} {Network structure of
  two-dimensional decaying isotropic turbulence},}\ }\href@noop {} {\bibfield
  {journal} {\bibinfo  {journal} {J. Fluid Mech.}\ }\textbf {\bibinfo {volume}
  {795}},\ \bibinfo {pages} {R2} (\bibinfo {year} {2016})}\BibitemShut
  {NoStop}%
\bibitem [{\citenamefont {Tezduyar}\ \emph {et~al.}(1992)\citenamefont
  {Tezduyar}, \citenamefont {Mittal}, \citenamefont {Ray},\ and\ \citenamefont
  {Shih}}]{tezduyar1992incompressible}%
  \BibitemOpen
  \bibfield  {author} {\bibinfo {author} {\bibnamefont {Tezduyar},
  \bibfnamefont {T.~E.}}, \bibinfo {author} {\bibnamefont {Mittal},
  \bibfnamefont {S.}}, \bibinfo {author} {\bibnamefont {Ray}, \bibfnamefont
  {S.}}, \ and\ \bibinfo {author} {\bibnamefont {Shih}, \bibfnamefont {R.}},\
  }\bibfield  {title} {\enquote {\bibinfo {title} {Incompressible flow
  computations with stabilized bilinear and linear equal-order-interpolation
  velocity-pressure elements},}\ }\href@noop {} {\bibfield  {journal} {\bibinfo
   {journal} {Comput. Meth. Appl. Mech. Engng}\ }\textbf {\bibinfo {volume}
  {95}},\ \bibinfo {pages} {221--242} (\bibinfo {year} {1992})}\BibitemShut
  {NoStop}%
\bibitem [{\citenamefont {Unni}\ \emph {et~al.}(2018)\citenamefont {Unni},
  \citenamefont {Krishnan}, \citenamefont {Manikandan}, \citenamefont {George},
  \citenamefont {Sujith}, \citenamefont {Marwan},\ and\ \citenamefont
  {Kurths}}]{unni2018emergence}%
  \BibitemOpen
  \bibfield  {author} {\bibinfo {author} {\bibnamefont {Unni}, \bibfnamefont
  {V.~R.}}, \bibinfo {author} {\bibnamefont {Krishnan}, \bibfnamefont {A.}},
  \bibinfo {author} {\bibnamefont {Manikandan}, \bibfnamefont {R.}}, \bibinfo
  {author} {\bibnamefont {George}, \bibfnamefont {N.}}, \bibinfo {author}
  {\bibnamefont {Sujith}, \bibfnamefont {R.}}, \bibinfo {author} {\bibnamefont
  {Marwan}, \bibfnamefont {N.}}, \ and\ \bibinfo {author} {\bibnamefont
  {Kurths}, \bibfnamefont {J.}},\ }\bibfield  {title} {\enquote {\bibinfo
  {title} {On the emergence of critical regions at the onset of thermoacoustic
  instability in a turbulent combustor},}\ }\href@noop {} {\bibfield  {journal}
  {\bibinfo  {journal} {Chaos}\ }\textbf {\bibinfo {volume} {28}},\ \bibinfo
  {pages} {063125} (\bibinfo {year} {2018})}\BibitemShut {NoStop}%
\bibitem [{\citenamefont {Williams}\ \emph {et~al.}(2017)\citenamefont
  {Williams}, \citenamefont {Kelley}, \citenamefont {Bersch}, \citenamefont
  {Br{\"o}ker}, \citenamefont {Campbell}, \citenamefont {Cunningham},
  \citenamefont {Denholm}, \citenamefont {Elber}, \citenamefont {Fearick},
  \citenamefont {Grammes} \emph {et~al.}}]{williams2017gnuplot}%
  \BibitemOpen
  \bibfield  {author} {\bibinfo {author} {\bibnamefont {Williams},
  \bibfnamefont {T.}}, \bibinfo {author} {\bibnamefont {Kelley}, \bibfnamefont
  {C.}}, \bibinfo {author} {\bibnamefont {Bersch}, \bibfnamefont {C.}},
  \bibinfo {author} {\bibnamefont {Br{\"o}ker}, \bibfnamefont {H.-B.}},
  \bibinfo {author} {\bibnamefont {Campbell}, \bibfnamefont {J.}}, \bibinfo
  {author} {\bibnamefont {Cunningham}, \bibfnamefont {R.}}, \bibinfo {author}
  {\bibnamefont {Denholm}, \bibfnamefont {D.}}, \bibinfo {author} {\bibnamefont
  {Elber}, \bibfnamefont {G.}}, \bibinfo {author} {\bibnamefont {Fearick},
  \bibfnamefont {R.}}, \bibinfo {author} {\bibnamefont {Grammes}, \bibfnamefont
  {C.}},  \emph {et~al.},\ }\bibfield  {title} {\enquote {\bibinfo {title}
  {gnuplot 5.2},}\ }\href@noop {} {\bibfield  {journal} {\bibinfo  {journal}
  {An interactive plotting program. Available online: http://www. gnuplot.
  info/docs\_5}\ }\textbf {\bibinfo {volume} {2}} (\bibinfo {year}
  {2017})}\BibitemShut {NoStop}%
\bibitem [{\citenamefont {Williamson}(1996)}]{williamson1996vortex}%
  \BibitemOpen
  \bibfield  {author} {\bibinfo {author} {\bibnamefont {Williamson},
  \bibfnamefont {C.}},\ }\bibfield  {title} {\enquote {\bibinfo {title} {Vortex
  dynamics in the cylinder wake},}\ }\href@noop {} {\bibfield  {journal}
  {\bibinfo  {journal} {Annu. Rev. Fluid Mech.}\ }\textbf {\bibinfo {volume}
  {28}},\ \bibinfo {pages} {477--539} (\bibinfo {year} {1996})}\BibitemShut
  {NoStop}%
\end{thebibliography}%
\end{document}